\documentclass[11pt,a4paper]{article}
\pdfoutput=1
\usepackage{jheppub}
\usepackage{amsmath,amscd}
\usepackage{amsfonts}
\usepackage{amssymb}
\usepackage{graphicx}
\usepackage{color}
\usepackage[normalem]{ulem}
\usepackage{hyperref}
\usepackage{enumerate}
\usepackage{mathtools}
\usepackage{slashed}

\newcommand{\CC}{\mathbb{C}} % Complessi
\newcommand{\RR}{\mathbb{R}} % Reali
\newcommand{\ZZ}{\mathbb{Z}} % Interi
\newcommand{\NN}{\mathbb{N}} % Naturali
\newcommand{\G}{\mathcal{G}}

\def\tr         {{\rm  tr}}
\def\cala         {{\cal A}}

\def\calb         {{\cal B}}

\def\calg         {{\cal G}}

\def\calh         {{\cal H}}

\def\calj         {{\cal J}}
\def\calk         {{\cal K}}
\def\call         {{\cal L}}
\def\calm         {{\cal M}}

\def\calo         {{\cal O}}

\def\calq         {{\cal Q}}
\def\calr         {{\cal R}}

\def\calu         {{\cal U}}

\def\calz         {{\cal Z}}

\def\be{\begin{equation}}
\def\ee{\end{equation}}
\def\bea{\begin{eqnarray}}
\def\eea{\end{eqnarray}}

\def\a{\alpha}
\def\b{\beta}

\def\g{\gamma}
\def\G{\Gamma}
\def\d{\delta}
\def\e{\epsilon}

\def\l{\lambda}
\def\L{\Lambda}
\def\k{\kappa}
\def\f{\phi}

\def\m{\mu}
\def\x{\xi}
\def\n{\nu}
\def\o{\omega}
\def\O{\Omega}
\def\p{\pi}

\def\r{\rho}

\def\s{\sigma}

\def\sF{{{ F}\!\!\!\!\hskip.8pt\hbox{\raise1pt\hbox{/}}\,}}
\def\som{{{ \omega}\!\!\!\!\hskip.8pt\hbox{\raise1pt\hbox{/}}\,}}
\def\sJ{{{\rm J}\!\!\!\!\hskip.8pt\hbox{\raise1pt\hbox{/}}\,}}

\def\F{\Phi}
\def\pa{\partial}

\def\to{\rightarrow}
\def\nonu{\nonumber \\{}}
\def\half{{1 \over 2}}

\usepackage{dsfont}

\title{Index and localization for type B superconformal mechanics on singular spaces}
\author[a]{Joris Raeymaekers,}
\author[a]{Paolo Rossi}
\author[a]{and Canberk \c{S}anl{\i}}
\affiliation[a]{CEICO, Institute of Physics of the Czech Academy of Sciences,\\  Na Slovance 2, 182 00 Prague 8, Czech Republic.}

\emailAdd{joris@fzu.cz}
\emailAdd{rossip@fzu.cz}
\emailAdd{sanli@fzu.cz}

\abstract{Type B superconformal quantum mechanical sigma models are of physical interest as they arise in the description of D-brane bound states forming an AdS$_2$ throat. In this work we discuss the applicability of   localization methods to compute  the superconformal index in these theories,  despite the fact that their  target spaces are generically singular.   Similar in spirit to recent  works on type A models, we propose to work on a suitably resolved target space to compute a regularized index. While this regularized index correctly captures the    actual index unambiguously in models of physical interest,  we do uncover a subtlety in more pathological examples. This occurs in situations where the supercharge is not essentially-selfadjoint, in which case the index becomes ambiguous and depends on the chosen selfadjoint extension. We also discuss the special class of models with K\"ahler target spaces, which can accommodate both type A and type B models, and show that the type B index is a particular limit of the type A index. For   Calabi-Yau cones, the   type B index coincides with  the Hilbert series of the unresolved space. 

}

\arxivnumber{}
\begin{document}
 \maketitle

  	\section{Introduction}
  Conformal symmetry in quantum mechanics \cite{deAlfaro:1976vlx} plays a role in a number of interesting physical systems ranging  from molecular mechanics to
  holographic quantum gravity. In the latter context, a supersymmetric extension of conformal symmetry is expected to govern the microphysics of  BPS black holes  which have an AdS$_2$ near horizon region. We refer  to  \cite{BrittoPacumio:1999ax,Fedoruk:2011aa} for reviews  on superconformal quantum mechanics and to \cite{Gibbons:1998fa,Claus:1998ts,Michelson:1999dx,Gaiotto:2004ij,Anninos:2013nra,Mirfendereski:2020rrk,Dorey:2022cfn}  for  an incomplete list of black hole-related  applications.

  An important quantity in such theories is the superconformal index, which carries    information about the short multiplet   spectrum and is expected to be computable using localization methods.
 In this work we will study this index for  
quantum mechanical sigma models. One distinguishes  models of `type A' and of `type B'. Type A models with $2N$-extended supersymmetry are constructed from multiplets with $N$ bosons and $2N$ fermions and arise from dimensional reduction of a 2D sigma model of type $(N,N)$. Type B models on the other hand employ multiplets with $N$ bosons and $N$ fermions and arise from  2D theories type $(0,N)$. Superconformal indices in type A models have been studied extensively by Dorey and collaborators \cite{ Barns-Graham:2018xdd,Dorey:2018klg, Dorey:2019kaf}. 
  The type A target spaces are K\"ahler cones and can, via embedding theorems,   often be realized as singular algebraic varieties \cite{Martelli:2006yb}. This allows for the application of powerful methods of algebraic geometry,  in particular  leading  \cite{Dorey:2018klg} to a natural proposal  for computing the index on an equivariant resolution of the target space. In this work,  we will instead study  type B models whose target spaces   are generically not K\"ahler and to which  the aforementioned algebraic geometry methods are typically not applicable. These models are nevertheless of interest for black hole physics. Indeed, our main motivating examples   are the conformal $N=4$B  models which arise  from the quiver quantum mechanics describing D-brane bound states in an AdS$_2$ decoupling limit \cite{Mirfendereski:2020rrk,Mirfendereski:2022omg}.

We will focus on conformal models with $N=2$ extended supersymmetry, the minimum for which a superconformal index can be defined. In order to compute the index using localization methods, one faces two obstacles. The first is that the target space is always noncompact and  that the number of BPS states contributing to the index is in fact infinite. In \cite{Raeymaekers:2024usy},  it was shown that type $2B$ models always posses an additional central $u(1)$ symmetry which can be used to define a refined index. The latter  is well defined as the number of BPS states with fixed $u(1)$ charge is finite. The noncompactness is not an issue for computing this refined index,  essentially because due to supersymmetric localization it is determined by the topological properties of the target space near the fixed locus of the $u(1)$ generator and is independent of the asymptotics.

In this work we address the second obstacle towards computing the index, namely  that the target space, and in particular  the fixed locus of the $U(1)$ action, is generically singular. We will therefore consider models on a resolved target space, which 
are not fully superconformal invariant yet preserve the subalgebra whose BPS states are counted by the index. 
Our approach is similar in spirit to the works \cite{Barns-Graham:2018xdd,Dorey:2019kaf} on the  type A  models, though phrased in the language of  differential rather than algebraic geometry. 
% These do not possess superconformal symmetry yet
The index of the resolved model, which can be computed reliably  using localization methods and does not depend on the details of the  resolution, provides a candidate expression for the superconformal index. We explicitly test this expectation in various examples where we can also directly compute the index in the conformal model. The latter requires a careful treatment of  boundary conditions at the singularity such that the superconformal generators are self-adjoint (see \cite{Reed:1975uy} for details on the theory of self-adjoint extensions).
We find that the regularized index correctly and unambiguously  captures the    actual index in models of physical interest where the supercharge annihilating the BPS states has the property of being `essentially-selfadjoint'. However,  we do uncover a subtlety in more pathological examples, in which the supercharge is not essentially-selfadjoint. In this situation the BPS spectrum and the index itself   become ambiguous and depend on the chosen self-adjoint extension.  In this case, the regularized index only captures a particular self-adjoint extension.
 A simple example where this caveat occurs is in two-dimensional target spaces with   a conical surplus singularity.

We also devote special attention to the special case of K\"ahler manifolds. These can serve as target spaces for both type A and type B models, and we will show that the type B index arises as a specific limit of the type A index. For targets which are Calabi-Yau cones, the results of \cite{Dorey:2019kaf} then allow us to  identify the type B index with an intrinsic geometric invariant of the unresolved space, namely its Hilbert series \cite{Forcella:2008bb}.

 \section{Review of $N=2 B$ superconformal sigma models}	
 In this work we will study supersymmetric quantum mechanical sigma models of type B,  in which  $N$-extended supersymmetry is realized on multiplets with $N$ real bosons and $N$ real fermions.  
 We refer to \cite{Smilga:2020nte} for a review of supersymmetric quantum mechanical sigma models, and to \cite{Michelson:1999zf, Mirfendereski:2020rrk} for more details on the conditions for superconformal symmetry.

Labelling the bosonic and fermionic worldline fields as $(x^A,\chi^A), A= 1,\ldots , D$, the general sigma-model Lagrangian is of the form
 \bea 
 L &=&  A_A \dot x^A  - {i } F_{AB} \chi^A \chi^B\nonu
  &&+   \half G_{AB} \dot x^A \dot x^B + {i }  G_{AB}\chi^A \hat \nabla_t \chi^B -  {1 \over 3} \pa_{[A} C_{BCD]}  \chi^A  \chi^B  \chi^C  \chi^D\label{L2gen}
 \eea	
 where the worldline covariant derivative on the fermions  is defined a
 \be
 \hat \nabla_t \chi^A := \dot \chi^A + \left( \G^A_{\ BC} + \half  C^A_{\ BC} \right) \dot x^B \chi^C.
 \ee
The various coupling functions appearing in the Lagrangian can be viewed as geometric structures on a $D$-dimensional target manifold $\calm$. The couplings $A_A$  play the role of a gauge potential for a background magnetic field strength $F_{AB}$, $G_{AB}$ defines a metric with  Christoffel symbols $\G^A_{\ BC}$,  and $C_{ABC}$ provides a   completely antisymmetric torsion tensor for a torsionful connection $\hat \nabla$ with connection coefficients  $\G^A_{\ BC} + \half  C^A_{\ BC}$.

Upon quantization, the canonical (anti-)commutation relations can be realized on  a Hilbert space consisting of spinors on $\calm$ with inner product
 \be
 (\F, \Psi) = \int_\calm  d^D x \sqrt{G} \F^\dagger \Psi.
 \ee
  In particular,  the quantum fermionic operators are realized as Dirac matrices
 \be
 \hat \chi^A = \G^A, \qquad \{ \G^A, \G^B \} = 2 G^{AB}.
\ee

As shown in \cite{Raeymaekers:2024usy}, the existence of a superconformal index requires at least $N=2$ extended supersymmetry, and we will restrict our attention to this minimal case in what follows. The conditions   for $N=2$B supersymmetry require the existence of an integrable complex structure $J^A_{\ \ B}$ such that\footnote{Slightly weaker conditions would in fact suffice \cite{Mirfendereski:2022omg}, but we will restrict to this subclass of models for simplicity.}
 \bea 
 F_{AC} J^C{}_B+F_{CB} J^C{}_A &=&0, \qquad	G_{AC} J^C{}_B+G_{CB} J^C{}_A=0.\label{N2conds}\\
\hat \nabla_A G_{B C}= \hat \nabla_A J^B_{\ \  C} &=&0.\label{Omcovconst}
 \eea
 In other words, the metric and field strength are hermitean with respect to $J^A_{\ \ B}$, while the metric and complex structure are covariantly constant with respect to the torsionful connection.
 The latter condition uniquely determines the torsion tensor to be given by
 \be
 C_{ABC} = - 3 J^D{}_A J^E{}_B J^F{}_C \nabla_{[D} J_{EF]} .\label{Cbism} 
 \ee
 and the corresponding torsionful connection $\hat \nabla$ is called the Bismut connection.  Models satisfying these conditions possess two selfadjoint\footnote{As we shall see below, for conformally invariant models the target space is singular and the definition of selfadjoint supercharges requires a careful consideration of boundary conditions at the singularity.} supercharges $Q_1, Q_2$ satisfying $\{ Q_i , Q_j\}= 2 H \d_{ij}, i = 1,2$. One of these, say $Q_2$, can be chosen   to be proportional to a  Dirac operator on the target space,
 \be  Q_2 = -  {i \over \sqrt{2}} \slashed{D}^{\rm tors}_{A},\label{Q2Dirac}
 \ee
 where $\slashed{D}^{\rm tors}_{A}$ indicates the Dirac operator with respect to the torsionful Bismut connection  
  and in the presence of the background magnetic potential $A_A$:
 \be
 \slashed{D}^{\rm tors}_{A} =     \G^A \left(\pa_A +{ 1\over 4} \left( \o_{ABC} - {1 \over 6} C_{ABC}\right) \G^B \G^C - i A_A \right).\label{DiractAtors}
  \ee

Choosing  complex coordinates $(z^m, \bar z^{\bar m})$ adapted to the complex structure, i.e.  such that
 \be 
J^m_{\ \ n} = i \d^m_n, \qquad   J^{\bar m}_{\ \ \bar n} = - i \d^{\bar m}_{\bar n}, \qquad  m = 1, \ldots, {D\over 2}\equiv d_\CC,
\ee
the nilpotent complex supercharge $\calq = (Q_1 + i Q_2)/2$, satisfying $\{ \calq, \calq^\dagger\} = H$, takes the form
 \be 
  \calq  =  {1\over \sqrt{2}} \G^{ m} \left(\pa_{ m} -i A_{ m}  + \half \o_{ m  n \bar p} \G^{ n} \G^{\bar p} 
+ {1\over 4} \pa_{ m} \ln{\sqrt{ G} \det \bar e\over \det   e} \right)\label{calqexpl}
 \ee
 We note that the last term simplifies if we choose a vielbein with real determinant, as we will do in what follows.

We also assume the existence of a $u(1)$ $R$-charge symmetry under which $\calq$ carries charge one.   This requires the existence of a real-holomorphic Killing vector $\r^A$ satisfying 
 \be   \call_\r G= \call_\r J  =0, \qquad i_\r F=0.\label{Rconds} \ee
 and the $R$-symmetry generator is then given by
\be
 R = i\left(   \r^A \left(\pa_A - i A_A  + {1 \over 4} \left( \o_{ABC} - \half C_{ABC} \right) \G^B \G^C \right)  + { 1\over 2} \nabla_A \r_B \G^A \G^B \right).
 \ee

We now want to impose that the model is in addition conformally invariant. This requires the  existence of a real-holomorphic  conformal Killing vector $\x$ satisfying: 
 \begin{align} 
 \call_\xi G_{AB} &= 2 G_{AB}, & \call_\xi J^A_{\ \  B} &=0\label{xixonds1}\\
 \call_\xi C_{ABC} &= 2 C_{ABC}, & i_\xi C &= 0.\label{xiCids}\\
  \r ^A &= - J^A_{\ \ B} \xi^B\, , &   \xi_A  &= \pa_A K.\label{Kxisq}
 \end{align}
 Here, the function $K$ is the generator of special conformal tranformations and is proportional to the norm of $\x$:
 \be 
  2 K =  \x^2.
  \ee
  It is possible to choose an  adapted radial coordinate such that \cite{Gibbons:1998xa}
  \be 
 \x =   r\pa_ r , \qquad K = {r^2\over 2},
 \ee
 and the metric takes the form of a cone over a base $\calb$
 \be 
 ds^2 = d r^2 + r^2 {ds}^2_\calb ,\label{conecoords}
 \ee 
and $\r$ is a Killing vector on $\calb$,
 \be 
 \call_\r G_\calb =0.
 \ee
 The target space of the conformal sigma model (\ref{conecoords}) therefore generically has  a curvature singularity at the tip of the cone $r=0$, unless the base metric $ds^2_\calb$ is the round metric on a $(D-1)$-sphere of unit radius. The base of the cone is said to possess a normal almost contact structure for which  ${ds}^2_\calb$ is an  adapted metric \cite{Boyer:2008era}. 
In the more familiar special case  where the Bismut torsion (\ref{Bismutcomps}) vanishes and the target space is K\"ahler, 
the base is  a Sasakian manifold. The relation (\ref{Kxisq}) between $\r$ and $\x$  implies that $\r$  then coincides with the Reeb vector field. 
 
 The superconformal charges $\calg_{\pm \half}$ are given by
 \be 
  \calg_{\pm \half} = e^{\mp K} \calq  e^{\pm K}\label{calgsim}
  \ee
and generate, together with $R$, the $su(1,1|1)$ superconformal algebra whose nonvanishing (anti-)commutation relations are given by
\begin{align}
\{\calg_{\pm \half},\calg_{\pm \half}^\dagger\}&=2L_0\pm R\label{BPS}, &
\{\calg_{\pm \half},\calg_{\mp \half}^\dagger\}&=2 L_{\pm 1}, &
\{\calg_{\a },\calg_{\b}\}&= 0\\
{} [L_0,\calg_{\pm \half}]&=\mp\frac{1}{2}\calg_{\pm \half}\label{rl}, & [L_{\mp 1},\calg_{\pm \half}]&=\pm \calg_{\mp \half}, & 
[R,\calg_{\pm \half}]&=\calg_{\pm \half}\\
{} [ L_m, L_n]&=(m-n) L_{m+n}.& & & &\label{algvir} 
\end{align} 
We will also introduce a fermion number operator defined as
\be 
F =  \half \G^{\underline m}\G_{\underline m}= {i \over 4} J_{AB} \G^A \G^B+ {d_\CC \over 2},\label{Fdef}
\ee
where underlined indices are frame indices.
In what follows we will use  the explicit representation for the Dirac matrices of  \cite{VanProeyen:1999ni}, namely
\be 
\G^{\underline m} = 2  \s^3\otimes \ldots  \s^3 \otimes \s^+ \otimes 1 \ldots \otimes 1, \qquad \G^{\underline{\bar  m}} = 2  \s^3\otimes \ldots  \s^3 \otimes \s^- \otimes 1 \ldots \otimes 1,\label{Gsexpl}
\ee
where the $\s^\pm $ entry is in the $m$-th position. The fermion number $F$ is represented as
\be 
F = \half ( 1+ \s_3) \otimes 1 \otimes \ldots \otimes 1 +\ldots+    1 \otimes \ldots \otimes 1  \otimes \half ( 1+ \s_3) \label{Fgen}
\ee
This shows that the fermion parity $(-1)^F $ coincides with the chirality operator
 \be 
 (-1)^F = (-1)^{d_\CC} \s_3 \otimes \ldots \otimes \s_3 = ( i)^{d_\CC} \G^{\underline 1} \ldots \G^{\underline D} \equiv   \G^{\underline{D+1}}.\label{Fpar}
 \ee
In \cite{Raeymaekers:2024usy} it was observed that a certain linear combination of $F$ and $R$ commutes with the $su(1,1|1)$ superconformal algebra, namely\footnote{ In writing (\ref{Jexpl})  we made use of the identity
 $ \r^A C_{ABC} = - 2 J_{AB} + 2 \nabla_A \r_B. 
 $}
 \be
    J =  F- R - {d_\CC \over 2} \equiv - i  L_\r - \r^A A_A \label{Jexpl},
 \ee
 where
 \be 
  L_\r= \r^A \left(\pa_A -i A_A + {1\over 4} \o_{ABC} \G^B\G^C\right)+ {1\over 4} \nabla_A \r_B \G^A \G^B\ 
  \ee
 is the spinorial Lie derivative with respect to the Killing vector $\r$. The $A$-dependent term in (\ref{Jexpl}) is necessary for covariance   with respect to abelian gauge transformations of $A$.

It is often the case that the  sigma model possesses an additional symmetry $J$ which commute with the superconformal algebra. This is  related to a real-holomorphic Killing vector $j$ possessing similar properties to the Reeb-like vector $\r$, namely \be \call_{j} G = \call_{j} J = \call_{j} K =0 , \qquad i_{j} F = d \m_j .\label{extrasymmconds}\ee 
Note that the last condition in (\ref{extrasymmconds}) on the gauge field  is weaker than what was required for   $\r$ in (\ref{Rconds}). It associates to every Killing vector a `moment map'  $\m_j$, determined up to an additive constant.
The corresponding symmetry generator acts on the Hilbert space  as a suitable covariantized spinorial Lie derivative
\be 
 J_j= - i  L_{j_I}- j^A A_A  - \m_j. \label{extrasymmgenr}
 \ee
In the special case of the Reeb-like vector $\r$ (\ref{Rconds}) implies that the moment map can be chosen to vanish,
\be \m_\r =0, \ee 
and   (\ref{extrasymmgenr}) reduces to (\ref{Jexpl}).  

\section{Regularized  index and  localization}\label{Secregprocedure} 
In this section we will define superconformal indices for  type B models and propose a regularization procedure in order to compute them using localization methods. We will also discuss a subtlety appearing in the case that the relevant supercharge is not essentially-selfadjoint.

\subsection{Superconformal indices}
Given a superconformal invariant model with $su(1,1|1)$ symmetry one could, in principle, consider the 
standard   `heat-kernel regularized' Witten index 
\be \tr (-1)^F e^{-\b H}\ee
receiving contributions from states annihilated by $\calq$ and $\calq^\dagger$. However, this
is  a highly subtle quantity since the Hamiltonian $H$ has a continuous spectrum extending down to zero because of the dilatation symmetry. However, in most models of physical interest, the operators $L_0$ and $R$ do have discrete spectra. We can therefore introduce  `conformally regularized' indices $\O_\pm$ 
 \be
	 \O_\pm = \tr (-1)^F e^{-\b \calh_\pm } .
	 \ee
   where
	 \be 
	 \calh_\pm = \{ \calg_{\pm \half}, \calg_{\pm \half}^\dagger \} = 2L_0\pm R.\label{calhpm}
	 \ee
   The indices $\O_+$ and $\O_-$ receive contributions from chiral primary states $|\chi^+\rangle$  resp. anti-chiral primary states $|\chi^-\rangle$ satisfying
	 \be 
 \calg_{\pm \half}	 |\chi^\pm \rangle =  \calg_{\pm \half}^\dagger	 |\chi^\pm \rangle =0.
 \ee
 However,
 as discussed in \cite{Raeymaekers:2024usy} the indices $\O_\pm$ are typically infinite\footnote{The reason for this is that, as  we shall see shortly, the target spaces of conformal sigma models are always noncompact, and the index gets contributions from an infinite number of lowest Landau level states in an effective magnetic field  (given by (\ref{gaugeshift})) below).}. 
 For this reason  we will rather study  refined  indices which keep track of the central $u(1)$ charge $J$: 
	 \be
	 \O_\pm [q] = \tr (-1)^F e^{-\b \calh_\pm } q^{ J},\label{Opmdef}
	 \ee
    These  have the usual `good' properties of an index, and are in particular independent of $\b$. The comments above imply that these indices diverge as $q\to 1$, and we will see that this is reflected in a pole in the $ \O_\pm [q] $. We should note that $J$ is in general not a canonically normalized $U(1)$ generator and its eigenvalues are not quantized\footnote{If the eigenvalues of $J$, normalized as in (\ref{Jexpl}), are integers, we can also define an alternative index weighted by a different parity $(-1)^R$ instead of $(-1)^F$. This index is then obtained (up to an overall factor) from $\O_\pm [q]$ by letting $q \to -q$. }. For example, they  vary continuously if we rescale the base metric  in (\ref{conecoords}). 
    %, which  rescales $\r$ and therefore $J$. 
     The index will therefore not be invariant under such deformations %, which should be viewed as changing the topology 
    (see \cite{Dorey:2019kaf} for a similar discussion for the type A index). It is of course possible to introduce a {\em model-dependent} rescaling of $J$ such that its eigenvalues become integer; upon doing so the index will then become a meromorphic function of $q$.

If the target space possesses additional Killing vectors $j_I$ which satisfy (\ref{extrasymmconds}) and commute with $\r$ and with each other, we can further refine the index to keep track of the associated charges $\calj_I$ given in (\ref{extrasymmgenr}) and consider
\be 
 \O_\pm [q, x_I] = \tr (-1)^F e^{-\b \calh_\pm } q^{ J} \prod_I x_I^{\calj_I}\label{Opmrefdef}.
\ee

 \subsection{Geometric interpretation} \label{sec:geom_interpretation}

It is useful to note that,	 due to (\ref{calgsim}), the conformal supercharges $ \calg_{\pm \half}$ and  `Hamiltonians' $\calh_\pm$ can be obtained from the Poincar\'e supercharge $\calq$ resp. the original  Hamiltonian $H$ by shifting the background gauge field as
	 \be 
	 A \to   \cala^\pm = A + \tilde A^\pm,\label{gaugeshift}
	 \ee
where
\be 
\tilde A^\pm =  \mp i (\pa - \bar \pa )K.\label{tildeA}
\ee
One shows that the holomorphic Killing vectors $\r$ and 
 (if present)  $j_I$ have nontrivial moment maps with respect to the additional gauge field $\tilde A^\pm$, namely
 \be 
 \tilde \m_\r^\pm = -  i_\r \tilde A^\pm = \mp 2K, \qquad  \tilde \m_I^\pm = -  i_{j_I} \tilde A^\pm.\label{mommapsAtilde}
 \ee
Splitting $\calg_{\pm 1/2}$ into real and imaginary parts, $\calg_{\pm 1/2} =(G_1^\pm + i G_2^\pm)/2$,  we can from  (\ref{Q2Dirac}) and   (\ref{DiractAtors}) express the 
  `Hamiltonians` $\calh_\pm$ in terms of Dirac operators as
 \be \label{eq:Dirac_op_wtorsion}
 \calh_\pm = \left( G_2^\pm \right)^2
 =  \half \left(\slashed{D}^{\rm tors}_{  \cala^\pm}\right)^2 .
 \ee
Thanks to the identities (\ref{mommapsAtilde}) one verifies that the $u(1)$ generators  $J = - i  L_\r^A -i_\r A$ 
and  $\calj_I = - i  L_{j_I}- i_jA - \m_I$ indeed commute with $\slashed{D}^{\rm tors}_{  \cala^\pm}$, despite being independent of $\tilde A^\pm $.

The refined superconformal index can therefore formally be identified with a  
`character-valued' Dirac index \cite{Witten:1983ux}
 as follows
 \bea 
\O_\pm [e^{i \theta}, e^{ i \n_I}] &=&   {\rm char_{g } \, Harm}^{+}\left(  i \slashed{D}^{\rm tors}_{\cala^\pm} \right)-  {\rm char_{g } \, Harm}^{-}\left(  i \slashed{D}^{\rm tors}_{\cala^\pm} \right)\\
&\equiv& {\rm ind_{g}}  \left( i \slashed{D}^{\rm tors}_{\cala^\pm} \right), \label{GindDirac}
\eea
where $g$ is the  group element
 \be 
g  = e^{i (\theta J  + \n^I \calj_I )},
\ee
 Here, the notation char$_g V$ denotes  the trace of the matrix representing the action of $g$ on  the vector space $V$, and ${\rm Harm}^{\pm}(  i \slashed{D}^{\rm tors}_{\cala^\pm} )$ are the spaces of positive res.  negative chirality harmonic spinors with respect to  $\slashed{D}^{\rm tors}_{\cala^\pm}$. 

 \subsection{Smoothening and localization} \label{sec:smooth_and_loc}
 
Despite the fact  that the superconformal index is   formally identified with an analytic Dirac index \cite{Ivanov:2010ki}, we can  not directly apply standard index theorems  to compute it since those are typically derived for compact, regular   geometries, and as we saw below (\ref{conecoords}), the target spaces of interest are noncompact and singular. In \cite{Raeymaekers:2024usy} it was argued that the noncompactness  is not an obstacle for applying index theorems to compute the refined index,  essentially since the latter is fully determined by local topological properties of    the target space in the vicinity of the vanishing locus of the $u(1)_J$ generator $\r$ (or,  allowing for further refinements, of the set of generators $\rho, j_I$).  

In this work  we aim to address the second  obstacle, which  is that the target space, and in particular said fixed locus, 
is generically singular.
%\change{\commentC{is there any analogy of this to higher dimensional SCFTS? Cyril Closset seems to think so. In particular he says this is related to the `square' of the supercharges used to localize the action (in our case anticommutator) being given by some combination of superconformal charges-which is indeed so in our case-. More concretely, I feel like in the localization we should address a point that despite we have $\{\calg,\calg^\dagger\} \sim su(1,1|1)$ we are still able to localize wrt $\calg$ (and show that this is special to 1d unlike higher dimensional cases) - naively maybe no such problem because `R' symmetry is broken-.}}{}
The direct computation of the BPS spectrum and the index    requires  a careful treatment of the boundary conditions at the singular locus so that the algebra generators are self-adjoint.

Nevertheless we will argue that the index can in many cases be reliably computed using localization methods. We will follow an approach similar in spirit to the treatment of type A models in \cite{Dorey:2019kaf}.  We will consider regularized sigma models on target spaces where the conical singularity is smoothened out such that the index can be computed using standard localization theorems. In type A models, which are K\"ahler cones and can mostly be realized as algebraic varieties \cite{Martelli:2006yb}, this was achieved in \cite{Dorey:2018klg,Dorey:2019kaf}  using the theory of equivariant resolutions in algebraic geometry. For generic type B models of interest, such techniques from algebraic geometry are typically not available and we will take a more differential-geometric approach to resolving the singularity.

We recall that the refined index makes use only of an $N=2$ subalgebra generated by $ \calg_{1/2}$ and $ \calg_{ 1/2}^\dagger$ ($ \calg_{-1/2}, \calg_{- 1/2}^\dagger$) and  central $u(1)$ generators $J$  and (if present) $\calj_I$. As we shall illustrate below, in many cases it is possible to   smoothen out the singularity at the tip of the cone while preserving this subalgebra\footnote{In particular, this requires the topology of the base manifold to be that of a sphere or a product of spheres.}. The resolved model  will of course not possess the  full $su(1,1|1) \oplus u(1)_J$ superconformal symmetry, which is only recovered in the singular limit. In practice, we will consider regular target space fields
$\{ A^{\rm reg}, G^{\rm reg}, J^{\rm reg}, K^{\rm reg}, \r^{\rm reg}, j_I^{\rm reg} \}$ (and corresponding Bismut torsion $C^{\rm reg}$) which satisfy the requirements (\ref{N2conds},\ref{Omcovconst},\ref{Rconds},\ref{extrasymmconds}) as well as
\be 
\call_{\r^{\rm reg}} K^{\rm reg} = \call_{j_I^{\rm reg}} K^{\rm reg} =0.\label{Kreginv}
\ee
 Thanks to these properties the resolved model possesses  $N=2$B supersymmetry with complex supercharge $\calq^{\rm reg}$ (see (\ref{calqexpl}))  and  commuting $u(1)$ charges
% \footnote{The condition (\ref{Kreginv}) could actually be weakened to $i_\r \pa \bar \pa K^{\rm reg} = d v_\r, i_{j_I} \pa \bar \pa K^{\rm reg} = d v_I$, at the cost of adding extra terms to the generators $J$ and $\calj_I$, namely 
%$\d J = \pm i i_\r (\pa - \bar \pa) K^{\rm reg} \mp 2 i v_\r $, 
%$\d \calj_I = \pm i i_{j_I} (\pa - \bar \pa) K^{\rm reg} \mp 2 i v_I $. These combinations vanish   when %(\ref{Kreginv}) is satisfied. \comment{In this more general setup the moment maps $v_I$ don't have to vanish on the %fixed locus and would enter the localization formulas. Shall we rather just not allow for this?}} 
$J^{\rm reg}, \calj_I^{\rm reg}$ (see (\ref{Jexpl},\ref{extrasymmgenr})).  
 Since we want to  modify the geometry  only near the tip of the cone, we will require that the geometric structures on the resolved space asymptotically approach their  conformal counterparts: 
\be
\F^{\rm reg} \xrightarrow[]{r_{\rm reg} \to \infty}\F, \qquad \F \in \{F,G, J,K,\r , j_I \}.
\ee
Here, $r_{\rm reg}$ stands for an appropriate radial coordinate in the resolved manifold approaching $r$ at large values. 
 We then use the function $K^{\rm reg}$ to define  similarity transformations  of the $N=2$ supercharges (and Hamiltonian)
\be
  \calg_\pm^{\rm reg} = e^{\mp K^{\rm reg} } \calq^{\rm reg}  e^{\pm K^{\rm reg} }, \qquad
  \calh_\pm^{\rm reg} = \{   G_\pm^{\rm reg},   (G_\pm^{\rm reg})^\dagger\}. \label{Gresolved}
  \ee
By construction, these approach the conformal supercharges
(and Hamiltonians) asymptotically, i.e.
   \be \calg_\pm^{\rm reg}\xrightarrow[]{r_{\rm reg} \to \infty}\calg_{\pm \half}, \qquad \calh_\pm^{\rm reg}\xrightarrow[]{r_{\rm reg} \to \infty}\calh_{\pm }.\ee
   
The above similarity transformation is  equivalent to shifting the gauge field   as in (\ref{gaugeshift}) by an extra piece
  \be 
 \tilde A^{\pm}_{\rm reg}= \mp i ( \pa - \bar \pa ) K^{\rm reg}.\label{tildeAreg}
 \ee
  Here, we should remark that, in order for (\ref{Gresolved}) to define a similarity transformation $K^{\rm reg}$ should, like $K$ in the conformal case, be a globally defined function. Therefore  the extra %field strength  $ \tilde F_\pm^{\rm reg}$  
  gauge field $\tilde A^{\pm}_{\rm reg}$ is a globally exact one-form and defines a connection on a trivial bundle. In particular, the flux   of its field strength 
  through any 2-cycle in the resolved manifold vanishes.
  One can show that the moment maps of the various Killing vectors with respect to the extra piece $\tilde A_\pm^{\rm reg}$ are, thanks to  holomorphicity and (\ref{Kreginv})
  \be 
 \tilde \m_{\r^{\rm reg}}^\pm = -  i_{\r^{\rm reg}} \tilde A^\pm_{\rm reg}, \qquad  \tilde \m_{j_I^{\rm reg}}^\pm = -  i_{j_I^{\rm reg}} \tilde A^\pm_{\rm reg}.\label{mommapsAtildereg}
 \ee

 In   regularized sigma models satisfying the above properties we can consider  refined $N=2$ indices
   \be
  \O_\pm^{\rm reg} [q, x_I] = \tr (-1)^F e^{-\b \calh_\pm^{\rm reg} } q^{ J^{\rm reg}} \prod_I x_I^{\calj_I^{\rm reg}}.
  \ee
These are good indices which are invariant under deformations of the target space metric and other fields which preserve the above geometric properties.  
Setting $q= e^{i \theta}, x_I = e^{i \n_I}$, the index localizes on the vanishing locus of the generators $\rho^{\rm reg}$, $j_I^{\rm reg}$.
% \be 
% j_{\theta, \n^I} = \theta \r^{\rm reg} + \n^I j_I^{\rm reg}.
% \ee
If this vanishing locus consists of isolated fixed points, we can compute the regularized index using the Atiyah-Bott formula  
\be
\Omega^{\rm reg}_{\pm}[e^{i\theta},e^{i\nu_I}] = 
\sum_{p:\rho,j_I = 0}\left. e^{ i\nu^{I}\mu_{I}^{\rm reg}}\prod_{a =1}^{d_\CC}\frac{1}{2 i \sin \frac{\theta l_{\rho,a} +  \nu^{I}l_{I,a}}{2} }\right|_p\label{ABformisolated}
\ee
where $l_{\bullet,a}(p)$ are the `exponents' of the torus action at the fixed point $p$, i.e. the charges  in the  decomposition of the action of the Killing vectors on the tangent space in irreducible representations.
%\change{Notice that in the resolved model we need to allow for a non-trivial moment map for the $\rho$-action }{} \commentP{right?}\comment{No, see below}. 
We give a concise review of how to derive this formula from the Atyah-Singer index theorem in Appendix \ref{App:Dirac_index}, as well as its generalization to non-isolated fixed points. In the latter case, the fixed point formula reads 
% \be
% \Omega_{\pm}^{\rm reg}[e^{i\theta},e^{i\nu_I}] = 
% \int_{\mathcal{M}^{T}}e^{F + i( \theta \mu_{\rho} + \nu^{I}\mu_{I})} 
% \prod_{r=1}^{\dim\mathcal{M}^{T}/2}\frac{x_{r}^{t}}{2\sinh(x_{r}^{t}/2)}
% \prod_{a=1}^{\mathrm{rank}N/2}\frac{1}{2\sinh(\frac{x_{a}^{n} +i \theta l_{\rho,a} +  i\nu^{I}l_{I,a}}{2})}
%\ee
\be
\Omega_{\pm}^{\rm reg}[e^{i\theta},e^{i\nu_I}] = 
\int_{\mathcal{M}^{T}}e^{c_1 (F^{\rm reg}) + i \nu^{I}\mu_{I}^{\rm reg}} 
\hat A (T \mathcal{M^T})
\prod_{a=1}^{\mathrm{rank}N/2}\frac{1}{2\sinh\frac{x_{a}^{n} +i \theta l_{\rho,a} +  i\nu^{I}l_{I,a}}{2}}\label{ABformnonisolated}
\ee
where $\cal M^T$ is the %\change{fixed point locus}{vanishing locus of  of $j_{\theta, \n^I}$},
vanishing locus of $\rho, j_I$, $c_1 (F) = - [{F/ 2\p}]$ is the first Chern class of the gauge bundle,  $N$ is the normal bundle to $T\cal M^T$, and  $x^n_a$ are the first Chern classes of $N$.
%and $x^t_r, x^n_a$ are the first Chern classes of $T\cal M^T$ and $N$ respectively.
% \begin{itemize}
% %\item  \comment{changed $F \to c_1 (F) = -[{F \over 2 \p}]$}
% %\item  \comment{replaced first product with $\hat A (\cal M^T)$, it's anyway 1 in all our examples 
% \item \comment{are there assumptions being made on the fixed locus being complex and the normal bundle splitting into line bundles diagonalizing the $u(1)$ action?}\commentP{No, it is true that the fixed point locus of a torus action is always of even codimension. Added a comment and a couple of footnotes in appendix.}
% \end{itemize}

One feature of the above formulas deserves further explanation, namely the fact that they are totally 
independent of the extra term $\tilde A_\pm^{\rm reg}$ in the  gauge field. This could contribute in two ways: firstly through the associated moment map, which due to (\ref{mommapsAtildereg}) reduces to 
%\change{$ - i_{j_{\theta, \n^I}} \tilde A_\pm^{\rm reg} $}{
$ - i_{j_I,\rho} \tilde A_\pm^{\rm reg} $. 
This however vanishes at the vanishing locus of %\change{$j_{\theta, \n^I}$}{
$j_I, \rho$. Secondly, it could contribute through its first Chern class $c_1 (\tilde F_\pm^{\rm reg})$, which is however trivial as we argued below (\ref{tildeAreg}).

\subsection{A possible caveat}
Though the above procedure produces a regularized index which is insensitive to the precise details of the regularization, a possible caveat is that the  index could in principle still jump discontinuously in the conformal limit. As  already stressed above, the direct computation  of the index in the conformal model
 requires a careful treatment of  boundary conditions at the singularity such that the superconformal generators are self-adjoint (see \cite{Reed:1975uy}  and Appendix \ref{app:selfadjoint_supercharge} for a review of self-adjoint extensions).
The index can jump discontinuously in the conformal limit if  the domain of the  Dirac  operator $\slashed{D}^{\rm tors}_{\cala^\pm}$ allows for zero modes which are more singular than those obtained as limits of regularized BPS states.
When $\slashed{D}^{\rm tors}_{\cala^\pm}$  is essentially-selfadjoint, this does not happen, since there is a unique self-adjoint extension which must coincide  with the limit of the regularized model. However, 
 when   it is not essentially-selfadjoint, we will see  that the BPS spectrum and the index itself become ambiguous and depend on the chosen self-adjoint extension.  In this case, the regularized index only captures a particular choice of self-adjoint extension. While in  physically relevant models, as far as we are aware, the Dirac operator $\slashed{D}^{\rm tors}_{\cala^\pm}$ appears to  be essentially-selfadjoint, we will identify a more pathological example (a conical excess in 2D) where it fails to be so.

For the standard Dirac operator  $\slashed{D}$  without torsion or gauge field on a compact manifold with a local conical singularity, the dependence of the zero-mode spectrum on the chosen selfadjoint extension was demonstrated by Chou in \cite{chou1985dirac}. 
%For the standard Dirac operator $\slashed{D}$ with vanishing torsion and background gauge field,  
%a well-known theorem of Chou \cite{Chou1989} states that  $\slashed{D}$  on a conical space 
It was shown there that  $\slashed{D}$ is essentially self-adjoint if and only if its eigenvalues $\l_i$ on the base
satisfy\footnote{It is a consequence of Lichnerowicz-Bochner-Weitzenböck formula that, if the base manifold has dimension greater than 1 and its scalar curvature is everywhere greater than that of the sphere, then \eqref{Choucrit} holds true.}
\be 
|\l_i^\slashed{D}| \geq \half \qquad \forall i.\label{Choucrit}
\ee
%\change{Our supercharge \eqref{eq:Dirac_op_wtorsion} is interpretable as the sum of the standard Dirac operator on the cone plus a 'torsion' term which does not involve derivatives. As long as this term is a bounded operator on the Hilbert space (in particular, regular at the tip of the cone), Chou's criterion directly extends to $\slashed{D}^{\mathrm{tors}}_{\cala^{\pm}}$. \commentP{This is actually a subtlety that we need to consider in more detail if we want to introduce a singular gauge field to the model.} }
We note that (\ref{Choucrit}) can always be violated by rescaling the base metric in (\ref{conecoords}). In our context,  the criterion (\ref{Choucrit}) will receive corrections, firstly  due to the effects of torsion and background gauge field in the Dirac operator $\slashed{D}^{\rm tors}_{\cala^\pm}$. A second modification is due the fact that (\ref{Choucrit}) arises from a {\em local} analysis near the singularity (as appropriate for compact manifolds), while we will have to impose suitable falloff conditions for normalizeablility at infinity. Here, due to the similarity transformation (\ref{calgsim}) (or equivalently the extra gauge field $\tilde A^\pm$ in (\ref{tildeA}), only selected states  from the local analysis will extend to normalizeable states on the infinite cone.
%that More drastic corrections to (\ref{Choucrit})  
%because of the presence of torsion and of a background gauge field, but also  because due to the  noncompactness we have to impose falloff conditions at infinity.
%will  also  since the relevant Dirac operator $\slashed{D}^{\rm tors}_{\cala^\pm}$ involves torsion and twist by a gauge bundle (even if $C=A=0$, we still have a twist by the extra gauge field $\tilde A^\pm$  which sensitively influences 
%the spectrum). 
Nevertheless, as we shall see in Section \ref{Sec2D}, for 2-dimensional target spaces (\ref{Choucrit}) is applicable to our situation without modification, while for higher dimensions  the criterion  can get drastically modified (see Section \ref{Sector}). %This theorem    gives a useful diagnostic for when the index is ambiguous for sigma models with vanishing torsion and background gauge field.  We observe that (\ref{Choucrit}) can always be violated by rescaling the base metric in (\ref{conecoords}). We will see however that in  the presence of  torsion and/or background magnetic field the condition (\ref{Choucrit}) on the eigenvalues of the standard Dirac operator is no longer a good criterion for essential-selfadjointness of $\slashed{D}^{\mathrm{tors}}_{\cala^{\pm}}$.}  the base metric. 
%\change{To summarize, in the case where the Dirac operator $\slashed{D}^{\mathrm{tors}}_{\cala^{\pm}}$  is not essentially-selfadjoint, i.e. when   Chou's criterion (\ref{Choucrit}) is violated,  there is a possible caveat  that the index might be ambiguous and depend on the chosen self-adjoint extension. 
% We will therefore pay special attention to this situation.
% We will see that a non-essentially  self-adjoint Dirac operator does not always lead to an ambiguity in the index; 
%   in fact in most case the ambiguity  resides only in the spectrum of nonzero modes and the index is still uniquely defined. However we will find one example where even the index is ambiguous, namely in the case of two-dimensional cones with an angular excess. In this case, the localization computation in the regularized model gives the index corresponding to one particular choice of self-adjoint extension. Note that this example is also pathological from physical point of view, as an angular excess would be produced by  the backreaction of a point-particle of  negative mass.}{}

\section{Examples: general 2D models}\label{Sec2D} 

In this section we will illustrate the computation of superconformal indices in the simplest class of examples, where the target space is 2-dimensional. We will explicitly compute the BPS spectrum in the conformal model, carefully treating boundary conditions at the singularity, and illustrate the procedure of smoothening out the singularity and computing the regularized index using localization formulas.  These models will furnish a concrete example  of the caveat described in the previous Section.
\subsection{Conformal models with 2D target space}
For 2-dimensional target spaces, using the fact that  
the base of the cone must possess a Killing vector (namely $\r$),  the canonical form of the metric  (\ref{conecoords}) is restricted to 
\be    
ds^2 = dr^2 + \a^2 r^2 d\f^2, \qquad \f \sim \f + 2 \p, \label{2Dconecan}
\ee
where $\a$ is a  real parameter which we will take to be positive in what follows. 
The  metric (\ref{2Dconecan}) is the flat metric on a cone with opening angle $2 \p \a $.
Except for $\a=1$ there is a conical singularity  in $r =0$. For $0<\a < 1$ we have a conical deficit, while for $\a > 1$ there is a conical excess. Note that for $\a = 1/N$ we are describing the $\CC / \ZZ_N$ orbifold.  

We introduce a complex  coordinate $z$ which takes values on the complex plane 
\be 
z = (\a r)^{1\over \a} e^{i \f}
\ee
in terms of which the metric becomes
\be
ds^2 = |z|^{2(\a-1)} dz d\bar z.\label{2Dmetrz}
\ee
One sees that the adapted complex structure with $J^z_{\ \ z} = - J^{\bar z}_{\ \ \bar z}= i$ satisfies the requisite properties (\ref{Rconds},\ref{xixonds1}), where
 the conformal Killing vector $\x$, the $u(1)_J$ Killing vector $\r$  and special conformal generator $K$ 
are related as in (\ref{Kxisq}) and  given by 
\be 
\x = {1 \over \a} ( z\pa_z + \bar z \pa_{\bar z} ), \qquad
\r = -{i \over \a} (z \pa_z - \bar z \pa_{\bar z} ),\qquad 
K= {|z|^{2\a} \over 2\a^2}.\label{rho2D}
\ee
 
Let us now work out the explicit form of the generators of the superconformal  algebra.  %$su(1,1|1) \oplus u(1)_J$. 
Choosing  the zweibein  
\be 
e^{\underline z} = |z|^{\a-1}dz,\qquad e^{\underline {\bar z}} = |z|^{\a-1}d \bar z;  \label{holframe}
\ee 
and using 
the   representation (\ref{Gsexpl}) for the Dirac matrices  
one finds find from (\ref{calqexpl}, \ref{calgsim}, \ref{Jexpl}) the following expressions  for  the basic $su(1,1|1)\oplus u(1)_J$ generators:
 \begin{align}  
 \calg_{\pm \half}  &= \sqrt{2} e^{\mp K} |z|^{3 (1-\a)\over 2 } \pa_z |z|^{ (\a-1)\over 2 }  e^{\pm K} \s_+, &  \calg_{\pm \half}^\dagger  &=- \sqrt{2} e^{\pm K} |z|^{3 (1-\a)\over 2 } \pa_{\bar z} |z|^{ (\a-1)\over 2 }   e^{\mp K} \s_- \label{gpmvan} \\
 J &= -{1\over \a}  \left(z \pa_z - \bar z \pa_{\bar z} + \half \s_3 \right) . & & % F &=\half \G^{\underline z}\G_{\underline z} = \half(1+\s_3), & (-1)^F &= i  \G^{\underline 1}\G^{\underline 2} = - \s_3
 \end{align}

 { Chiral primary} states of the superconformal algebra are by definition annihilated by $\calg_{1/2}$ and $\calg_{1/2}^\dagger$. Using the above expressions, one finds the following complete set of    normalizeable  solutions   
 \be 
 |\chi^+_n \rangle = \bar z^n |z|^{1-\a\over 2} e^{ - K} | \downarrow \rangle, \qquad {\rm for\ }  n\in\ZZ, 
 n > -{\a + 1\over 2} .\label{cp2d}
 \ee
Similarly, a complete set of normalizeable  {anti-chiral primary} solutions, which are annihilated by $\calg_{-1/2}$ and $\calg_{-1/2}^\dagger$, is furnished by
 \be 
 |\chi^-_n \rangle = z^n |z|^{1-\a\over 2} e^{ - K} | \uparrow \rangle, \qquad {\rm for\ } n\in\ZZ, n > -{\a + 1\over 2}  .\label{acp2d}
 \ee
{Let us remark that the states $|\chi^{\pm}_n\rangle $, for $n\in\NN$ are at the same time normalizable in the measure $d^2z |z|^{2(\a -1)}$ and regular at $|z|\to 0$ only when the cone parameter is in the conical defect range $\a\in (0,1]$. In the conical surplus regime $\a >1$ some of the wavefunctions diverge at the singularity while remaining square-integrable, moreover some states with $n<0$ also become normalizable. 
This qualitatively different behavior for $\a >1$ is related to the Dirac operator ceasing to be essentially-selfadjoint in this regime. Indeed, let us examine Chou's criterion (\ref{Choucrit}) in these models.\footnote{Let us stress that our operator $G_2^{\pm}$ is not exactly the standard Dirac operator, due to the presence of the effective potential $(\pa -\bar\pa )K$. This term affects the behavior of the BPS states at $r\to \infty$, while \eqref{Choucrit} relates to the behavior near $r\to 0$. Thus Chou's criterion gives a \emph{necessary} (but not sufficient) condition for the existence of extra normalizable BPS states, which would lead to ambiguities in the spectrum. We can anticipate that in the 2D case these extra states are indeed normalizable (they have the correct falloff behavior at infinity) and Chou's criterion gives a sensible result.} %\comment{Argument that the effective $F =\pa \bar \pa K$ does not play a role?} \commentP{Chou tells us that there can be an ambiguity, below we confirm that there is (by direct check in appendix). No need to be more detailed here in my opinion} 
To split the 2D Dirac operator in radial and angular parts, we use the zweibein $(dr , \a r d\f)$ in the coordinates (\ref{2Dconecan}). It is then straightforward to see that the angular part becomes $\slashed{D}_{S^1} = \a^{-1} \pa_\f$ acting on antiperiodic functions on the unit circle. The eigenvalues are 
\be
\l_n = {n+ \half \over \a },\label{DspecS1}
\ee
and Chou's criterion (\ref{Choucrit}) is violated precisely for $\a>1$. 
As we shall see in more detail below, in this regime  there  is an ambiguity in the BPS spectrum which depends on the chosen self-adjoint extension. For the remainder of this subsection, let us therefore focus on the  conical defect regime where $\a\in(0,1]$.

To compute the index in this regime, we note that the (anti-)chiral primaries  carry $u(1)_J$ quantum numbers    \begin{align} \label{jaction2d}
    J  |\chi^+_n \rangle & = {1\over \a}\left( n +\half \right)|\chi^+_n \rangle, &  J  |\chi^-_n \rangle & = -{1\over \a}\left( n +\half \right)|  \chi^-_n \rangle 
     	\end{align}
      Evaluating the   indices (\ref{Opmdef}) using that the fermion parity is $(-1)^F= -\s_3$ (see (\ref{Fpar})), we  find
     	\bea \label{eq:index_conica_deficit_bruteforce}
     	 \O_\pm [q ] 
     	 &=&\left[ {q^{1 \over 2  \a} \over 1-q^{1\over \a}} \right]_\pm
     	\eea
       where $[\ldots ]_+$ and $[\ldots ]_-$ denote the Laurent series  in positive resp. negative powers
  of $q$.
     	It's satisfying that the two indices, which can be seen as  different `conformal  regularizations' of the same object $\tr (-1)^F  q^{2J}$, correspond to  different Laurent expansions of the same meromorphic function.

Before discussing the conical surplus regime, let us remark that our analysis of models with 2D target spaces has not been completely   general, since we have not allowed for a background gauge field $A_A$ in the Lagrangian (\ref{L2gen}). It's straightforward to see that the conditions (\ref{N2conds},\ref{Rconds}) require this to be  pure gauge away from the singularity.
However, at the singularity we should for   completeness  allow  the   field strength to have a delta-function singularity. That is, we allow for  a gauge potential of the form 
\be 
A = i b \left( {d\bar z\over \bar z} - {dz \over z}\right) = i d \ln \left( {\bar z \over z } \right)^b \label{Asing}
\ee 
for some real constant $b$. The gauge field is locally pure gauge but not globally since the gauge parameter $\L = i b\ln (\bar z/z)$ is not single-valued for generic $b$. Under such an (improper) gauge  transformation the (anti)-chiral primary wavefunctions transform as fields of charge one, i.e. the  expressions (\ref{cp2d},\ref{acp2d}) are multiplied by a phase factor $e^{i \L}$. 
Furthermore, the $u(1)_J$ generator picks up an extra term (see (\ref{Jexpl})) $ - i \r^A A_A = 2 b /\a $. The net effect is  that the 
$J$-eigenvalues of the BPS states remain unchanged, as they of course should, being gauge-invariant observables. The indices (\ref{eq:index_conica_deficit_bruteforce}) are therefore unmodified.

\subsection{BPS spectra on conical surplus geometry} 
Let us consider again the BPS states $|\chi^\pm_n\rangle$ and explore in more detail their behavior around the conical singularity $|z|\to 0$. As already remarked, when $\alpha\in(0,1]$ the only normalizable states have $n\in\{0,1,2,\cdots\}$ and they are regular at the singularity. They define unambiguously the BPS spectrum in the range of conical defects, giving rise to the index \eqref{eq:index_conica_deficit_bruteforce}. As explained in detail in Appendix \ref{app:selfadjoint_supercharge}, in this regime the hermitian combinations of supercharges 
\be
G_1^{\pm}=\mathcal{G}_{\pm\frac{1}{2}}+\mathcal{G}_{\pm\frac{1}{2}}^{\dagger},\qquad G_2^\pm =i(\mathcal{G}_{\pm\frac{1}{2}}-\mathcal{G}_{\pm\frac{1}{2}}^{\dagger})\propto\slashed{D}^{\mathrm{tors}}_{\cala^\pm}
\ee
are \emph{essentially self-adjoint}.\footnote{As shown in Appendix \ref{app:selfadjoint_supercharge}, the  charges $G_1^{\pm}$ and $G_2^{\pm}$ are unitarily related so we only need to study the spectrum of one of the two.} 
This in our context implies, roughly speaking, that there is a unique appropriate boundary condition for the wavefunctions which is automatically respected by their eigenstates. So the BPS spectrum is uniquely determined in terms of the square-integrable solutions of the equations $G_1^{\pm}|\chi^{\pm}\rangle=0$, with no need of imposing extra conditions. 

When $\alpha>1$, a little inspection shows that the states with $-\frac{\alpha+1}{2}<n<\frac{\alpha-1}{2}$ blow up near the singularity, still being normalizable, while the states with $n\geq\frac{\alpha-1}{2}$ are both regular and normalizable. Let us call the former “negative” and the latter “positive” BPS states. This has to do with the fact that $G_1^{\pm}$ are not essentially self-adjoint in the regime of conical surplus (again, for details see Appendix \ref{app:selfadjoint_supercharge}). This implies that %, roughly speaking, 
there are different choices of appropriate boundary conditions for the wavefunctions around the singularity. Each  of these choices %\change{is physical} {leads to a consistent model with  different physics} 
leads to a consistent model with  different physics and, when imposed on the solutions of the BPS equation $G_1^{\pm}|\chi^{\pm}\rangle=0$, determines which of them has to be kept or discarded from the BPS spectrum. In particular, different choices correspond to keeping only specific linear combinations of the negative BPS states and discard the others, while positive BPS states always belong to the complete spectrum. See Figure \ref{fig:neg_pos_BPS_states}.

\begin{figure}
    \centering
    \includegraphics[width=0.5\linewidth]{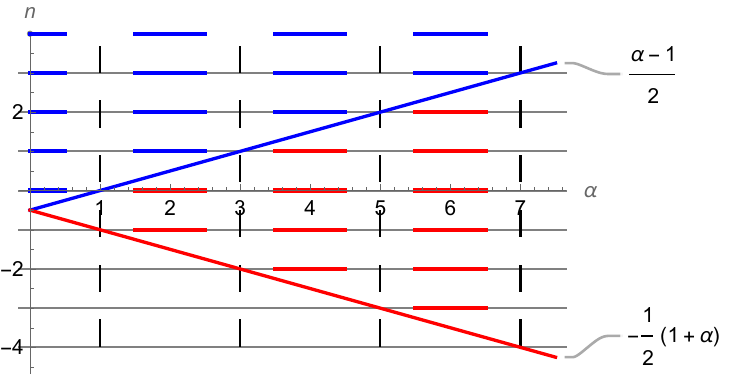}
    \caption{Structure of BPS states $|\chi_{n}^{\pm}\rangle$ arranged with respect to their “angular” quantum number $n$, at various values of the cone parameter $\alpha$. Blue horizontal bars denote positive BPS states while red horizontal bars denote negative BPS states. When $\alpha<1$, all states are positive and the spectrum is uniquely determined. When $\alpha\in(2\kappa-1,2\kappa+1)$ for some integer $\kappa$, states with $n\geq\kappa$ are positive and states with $-\kappa\leq n\leq\kappa-1$ are negative.}
    \label{fig:neg_pos_BPS_states}
\end{figure}

In the range of conical surplus, choosing one self-adjoint boundary condition or another results in changes in the BPS spectrum and in turn in the superconformal index. For example, \eqref{eq:index_conica_deficit_bruteforce} is the unique result in the range of essential self-adjointness and it depends very mildly on the cone parameter $\alpha$: as commented below (\ref{Opmdef}), sending $q \to q^{2 \a}$  
renormalizes the quantum charge $J$ to have integer eigenvalues and completely absorbs this dependence. This resonates with the topological property of the index, and suggests that when the charges are essentially self-adjoint one can safely consider the singular geometry as a continuous limit of its smooth regularization. Instead, when $\alpha\in(2\kappa-1,2\kappa+1)$ for some $\k \in \NN_{>0}$, to reproduce \eqref{eq:index_conica_deficit_bruteforce} we need to make a specific choice of self-adjoint boundary conditions. One choice, corresponding to the “minimal” BPS spectrum, is to keep only the positive BPS states with $n\geq\kappa$: the index in this case turns out to be
\be
\Omega_{\pm}^{\mathrm{min}}[q]= q^{\pm \frac{\kappa}{\alpha}}\Omega_{\pm}[q].
\ee
Another example corresponds to the “maximal” BPS spectrum, where we keep all normalizable states with $n\geq-\kappa$: the index would be
\be
\Omega_{\pm}^{\mathrm{max}}[q]= q^{\mp \frac{\kappa}{\alpha}}\Omega_{\pm}[q].
\ee
%\change{Even with the normalization $c_J=\a $, }{T}
The ratio between these indices depends inequivocally on the \emph{discrete} quantity $\kappa$ (which we then interpret in some sense as “topological”). 

Simple choices of boundary conditions like the ones 
above make us keep negative BPS states up to some quantum number $n\geq M$. In these cases the index 
changes with a phase shift %\change{$q^{\pm 2M \frac{c_J} {a}} \O_\pm[q]$} {
$q^{\pm  \frac{M}{a}} \O_\pm[q]$, which 
could be reabsorbed as an ordering constant in the 
operator %\change{$J\mapsto J \pm 2 M \frac{c_J}{a} $}
{$J\mapsto J \pm \frac{M}{a} $}. Different boundary 
conditions $\calu$ can produce more complicated changes
in the index, which cannot be interpreted in such a way: 
generically we have
\be
\O^{\calu}_\pm [q] = \mathrm{Tr}|_{\mathrm{neg.BPS}(\calu)} (-1)^F q^{J} + \O^{\mathrm{min}}_\pm [q]
\ee
where the first term comes from the negative BPS states and depends on $\calu$, while the second term is the universal contribution from the positive BPS states.

% \textbf{TO DO}: relate ambiguous phase dependence on $\kappa$ with “spectral asymmetry” of the Dirac operator? Related to this, the phase can be interpreted as an "ordering constant" for the quantum charge $J$.

% \change{\textbf{TO DO}: relate eigenvalues of transverse dirac op to the conical parameter, and meet Chou's criterion. Justify that until the “twist/torsion/superpotential” part of our Dirac operator is a bounded operator, the same criterion for self-adjointness for the bare Dirac operator of Chou holds. This assumes regularity of the “twist/torsion/superpotential” part at the singularity!}{\commentP{Added section below and comment near to Chou's crit.}}

  \subsection{Resolved models and localization formula}
   In Section \ref{Secregprocedure} we proposed a method for computing the superconformal index by working on a smoothened  target space. 
   Here we will explicitly  illustrate such a regularization  for the above general models with 2D targets and discuss  a  caveat that can appear  when the supercharge is not essentially-selfadjoint.  The simplicity of these examples will allow us to compute the regularized BPS wavefunctions explicitly,  and discuss their convergence to  BPS wavefunctions derived earlier in the conformal limit. We also show how the index can be computed painlessly in the resolved model using the localization formula (\ref{ABformisolated}).

 To regularize our models,  we  replace target space metric of the conformal sigma model (\ref{2Dmetrz})  regular metric of the form
     \be 
    ds^2_{\rm reg} = f(|z|)^2 dz d\bar z,\label{metrreg2D}
    \ee 
     where $f$ is a smooth function such that  $f(R) =1$ in a neighborhood   of $R=0$, while $f(R) \sim R^{\a-1}$ for $R\to \infty$ 
     %\commentC{interpretation of $\alpha$ in the resolved metric is a bit confusing to me. There should be a topological meaning of $\alpha$ as an invariant in the K\"ahler geometry (of resolved space) -otherwise the index depending on $\alpha$ is contradictory with general arguments-. It would be nice if we can rewrite this resolved metrics in a way where the actual resolution parameter/function is independent from $\alpha$.}. 
     The index will of course not depend on the precise form of $f$. For concreteness, one can consider the family of functions parametrized by a (small) parameter $\e$ (see Figure \ref{figreg}),
     \begin{figure}
    \centering
    \includegraphics[height=120pt]{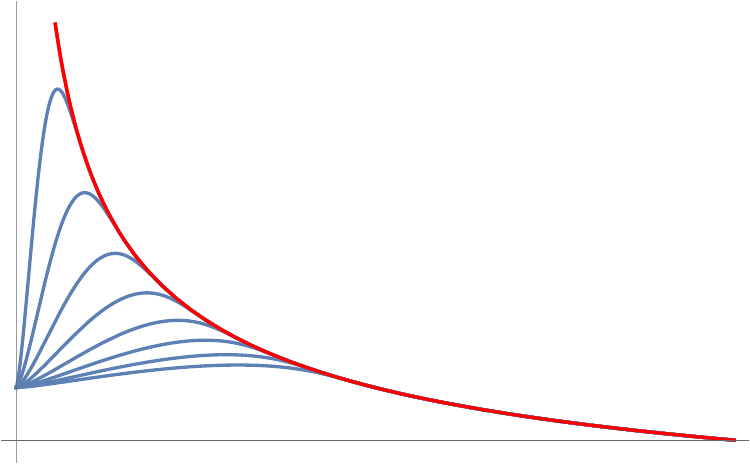}
    \caption{The regulating  functions $f_\e (R)$ of eq. (\ref{bump}) for various values of $\e$. The red curve is $f_{\rm conf} = R^{\a-1}$.}
    \label{figreg}
\end{figure}
     \be 
     f_\e (R) = B_\e (R) + \left(1- B_\e (R) \right)R^{\a-1}, \qquad B_\e (R) =\left\{ \begin{array}{ll}
     e^{{1\over \e^2} - {1\over \e^2 - R^2}} & 0\leq R<\e\\ 0 & R\geq \e\end{array}\right.\label{bump}
     \ee
  so that for $\e\to 0$ the singular conformal model with $f_{\rm conf} (R) = R^{\a-1}$ is recovered.   We note that,  $f_\e$ converges pointwise to $f_{\rm conf}$ on $\RR_{>0}$ but (of necessity) not uniformly. This will however be sufficient for the BPS wavefunctions to converge in $L^2$ norm as we shall see shortly. 
  %\change{ \commentP{The function $f(R)=R^{\a-1}$ is only defined on $R\in (0,\infty)$. So it is correct to say that the sequence $f_{\epsilon}(R)$ converges \textit{pointwise} to $f(R)$ on $R\in(0,\infty)$, as $R=0$ is out of the domain. It does not, however, converge \textit{uniformly} because there is no global estimate for the difference $|f_{\epsilon}(R)-f(R)|$ on $R>0$. This does not matter, as we only care about the states $|\chi_n^{\pm}\rangle_\epsilon$ converging to $|\chi_n^{\pm}\rangle$ in $L^2(\mathbb{R}_{>0},R^{2\a-1}dR)$, which they do (the precise measure is irrelevant for this).}}{}
  Since the singular gauge  field (\ref{Asing}) had vanishing field strength everywhere except at the tip of the cone, we can take the regularized gauge field to simply vanish,
\be 
A^{\rm reg} = F^{\rm reg} =0.
\ee

     These regularized models satisfy the conditions for $N=2$B supersymmetry commuting with a $u(1)$ symmetry. Indeed, the metric is Hermitean with respect to the complex structure associated to the coordinates $(z,\bar z)$, i.e. 
     $(J^{\rm reg})^z_{\ z} =i = - (J^{\rm reg})^{\bar z}_{\ \bar z}$,
      and $J^{\rm reg}$  is covariantly constant with respect to the Levi-Civita connection. 
    Asymptotically, for $|z|\to \infty$, we approach the conformal model which possesses a real-holomorphic Killing vector $\r$ given in (\ref{rho2D}). This vector extends to a real-holomorphic Killing vector on the full regularized target space, so we can take
     \be
    \r^{\rm reg}  = -{i \over \a} (z \pa_z - \bar z \pa_{\bar z} )
     \ee 
   and check that
     \be 
     \call_{\r^{\rm reg}} G^{\rm reg} = \call_{\r^{\rm reg}} J^{\rm reg} =0
     \ee
    We will also pick a smooth function $K^{\rm reg}$ of $|z|$ with asymptotics
     \be K^{\rm reg} (|z|) \to K = {|z|^{2\a}\over 2 \a^2} \qquad {\rm for \ } |z|\to \infty ,
     \ee
     which satisfies $\call_{\r^{\rm reg}} K^{\rm reg} =0$.
   The resolved model therefore satisfies all the requisite properties for defining a regularized index  
\be    \O^{\rm reg}_\pm [q ] = \tr (-1)^F e^{ - \b \calh_\pm^{\rm reg}}q^{ J^{\rm reg}}. \ee

Let us first compute $\O^{\rm reg}_\pm $ by brute force  by constructing the  BPS states explicitly.     Taking the zweibein
     \be 
     e^{\underline z}_{\rm reg} = f(|z|) dz,\qquad e^{\underline{\bar  z}}_{\rm reg} = f(|z|) d\bar z
     \ee
     one finds that the supercharges  take the form  
      \be
     \calq_{\rm reg} =  \sqrt{2} f^{-{3\over 2}} \pa_z f^{{1\over 2}}  \s_+, \qquad 
      \calq^\dagger_{\rm reg} = -  \sqrt{2} f^{-{3\over 2}} \pa_{\bar z} f^{{1\over 2}}  \s_- .\label{calqregd2}
     \ee
     The modified supercharges $\calg^{\rm reg}_\pm$  are obtained  by a similarity transformation % and which asymptotically approach $\calg_{\pm \half}$,
   \linebreak  $ \calg^{\rm reg}_\pm = e^{\mp K^{\rm reg} } \calq^{\rm reg} e^{\pm K^{\rm reg}}$,
     and asymptotically approach $\calg_{\pm \half}$.
     The $U(1)$ generator $J^{\rm reg}$ still takes the form
     \be 
     J^{\rm reg} = - i  L_{\r^{\rm reg}} = -{1\over \a}  \left(z \pa_z - \bar z \pa_{\bar z} + \half \s_3 \right) .
     \ee
     It's straightforward to check that $ J^{\rm reg}$ indeed commutes with $\calg^{\rm reg}_\pm$. 
     
     From these expressions one finds a basis of normalizeable BPS states annihilated by $\calg^{\rm reg}_+, (\calg^{\rm reg}_+)^\dagger$ resp. $\calg^{\rm reg}_-, (\calg^{\rm reg}_-)^\dagger$:
    \begin{align}
     |\chi^{{\rm reg}, +}_n \rangle & = \bar z^n f^{-\half}  e^{ - \calk }|\downarrow \rangle& | \chi^{{\rm reg}, -}_n \rangle & =  z^n   f^{-\half}  e^{ - \calk }|\uparrow \rangle & {\rm for}\  n \in \NN
     \end{align}
     Here we observe that normalizeability restricts $n$ to be positive independent of $\a$, which contrasts  with the conformal models for $\a>1$. Taking for example the  regulating function to be $f_\e$  given in (\ref{bump}), one sees that the regularized wavefunctions converge to the $n\geq 0$ conformal wavefunctions (\ref{cp2d},\ref{acp2d}) in $L^2$-norm,  in  the sense that
     \be 
     || g_\e^{1/4} |\chi^{\e, \pm}_n \rangle - g^{1/4} |\chi^{ \pm}_n \rangle || \to 0 \qquad  {\rm for}\  n \in \NN
     \ee
     %\comment{For singular gauge field $b \neq 0$ we have to remove a phase $e^{i\L}$ in the conformal wavefunctions, but let's not confuse matters by drawing attention to this.}
in  $L^2 (\CC \backslash \{0\}, d^2 z )$. This illustrates that in the case when $\a >1$ and the limiting supercharge is not essentially self-adjoint,  the regularized model captures only one of the  many possible self-adjoint extensions.  

  From the $u(1)_{J^{\rm reg}}$ quantum numbers
     \begin{align}
    J^{\rm reg}  |\chi^{{\rm reg}, +}_n \rangle & = {1\over \a}\left( n +\half \right)|  \chi^{{\rm reg}, +}_n \rangle, &  J^{\rm reg}  | \chi^{{\rm reg}, -}_n \rangle & = -{1\over \a}\left( n +\half \right)|  \chi_n^{{\rm reg}, -} \rangle
     	\end{align}
     	we find for the resolved indices
     	\bea
     	 \O^{\rm reg}_\pm [q ] &=& \tr (-1)^F e^{ - \b \calh_\pm^{\rm reg}}q^{ J^{\rm reg}}\\
     	 &=&\left[ {q^{1 \over 2  \a} \over 1-q^{1\over \a}} \right]_\pm\label{omreg2d}
     	\eea
   
As argued in Section \ref{Secregprocedure}, the regularized index can also be computed efficiently using standard localization formulae. 
The group action generated by $\r^{\rm reg}$ has an isolated fixed point in $|z|=0$, so that the Atiyah-Bott formula (\ref{ABformisolated}) is  applicable. The $u(1)_J$ action on the tangent vector $\pa_z$ at the fixed point is given by
\be
\call_{\r^{\rm reg}} \pa_z = {i \over \a} \pa_z,
\ee
and therefore the corresponding charge or exponent is $l= 1/\a$.
Substituting in (\ref{ABformisolated}) we obtain for the regularized index
\be 
\O^{\rm reg}[q] =  {q^{1\over 2\a } \over 1- q^{ 1  \over \a}}.\label{fixedpindex2D}
\ee 
So far the localization computation was not sensitive to whether we want to count chiral primary degeneracies (i.e.  $\O^{\rm reg}_+[q]$) or anti-chiral ones (i.e.  $\O^{\rm reg}_-[q]$). However, due to the pole in $q=1$, the   function (\ref{fixedpindex2D})  possesses   two possible Laurent expansions. To figure out which series belongs to which index, we use the physical input that the spectrum of the model is expected to be comprised of unitary, lowest $L_0$ weight representations of the conformal group $SL(2,\RR)$. This means that all $L_0$-eigenvalues are positive and, using (\ref{calhpm},\ref{Jexpl})  chiral primaries have $J$-eigenvalues which are bounded below, while anti-chiral primaries have $J$-eigenvalues which are bounded above. Therefore we have to pick the corresponding Laurent series which are in agreement with the brute force calculation  (\ref{omreg2d}).

In conclusion, we have seen that the regularized index, which can be computed easily from  localization formulae, correctly captures the actual superconformal index as long as the supercharge is essentially-selfadjoint. However,  the index becomes ambiguous for non-essentially-selfadjoint supercharges, and in this case the regularized index captures one of the many possible selfadjoint extensions.

\section{Examples with torsion}  \label{Sector}

The two-dimensional sigma models considered so far were somewhat special, in that the torsion tensor $C_{ABC}$ necessarily vanishes and the target is  actually K\"ahler. To illustrate  our method for the more general complex manifolds with torsion  we consider the following generalization of the 2D metric (\ref{2Dmetrz}). For general even dimension $D= 2 d_\CC$, we introduce a family of Hermitean metrics
\be 
ds^2 = R^{2(\a-1)} d z^m d\bar z^{\bar m}, \qquad R^2 = z^m \bar z^{\bar m},\qquad m = 1, \ldots, {d_\CC}\label{metrtorsex}
\ee
where $z^i$ are complex coordinates adapted to the complex structure. The $N=2$B supersymmetric sigma model  Lagrangian includes  nonvanishing Bismut torsion, whose components are given by the general formula \cite{Smilga:2020nte}
\be 
C_{mnp}= 0, \qquad C_{mn \bar p}= -2 \pa_{[m}G_{n]\bar p}.\label{Bismutcomps}
\ee
We should note that the metrics with $\a \to - \a$ are isometric and related by the `inversion' transformation
\be 
\tilde z^m = {\bar z^{\bar m} \over R^2}.\label{inversion}
\ee
We will therefore restrict our attention to the case\footnote{This does constitute a slight loss of generality, since the inversion (\ref{inversion}) is not a holomorphic transformation for $D>2$. The models  arising from an a priori negative $\a$ have a metric of the form (\ref{metrtorsex}) with $\a \to |\a|$ but  an inequivalent complex structure
$ \tilde J = i d\tilde z^m \pa_{\tilde z^m}- i d\bar{\tilde z}^{\bar m} \pa_{\bar{\tilde z}^{\bar m}}$ and correspondingly a Bismut torsion different  from (\ref{Bismutcomps}). We will not consider the models with this alternative complex structure here.}$\a>0$.
%the corresponding sigma models are {\em not} equivalent for $D>2$. This is because in that case the coordinate transformation is not holomorphic, and therefore the models have different complex structures and torsion tensors. 
%Again, sending $R\to R^{-1}$ replaces $\a$ by $-\a$ and we can assume $\a$ to be positive. 
%\change{\comment{Puzzle: both $\a=1$ and $\a=-1$ describe flat space, but for the former the torsion (see below) vanishes, while for the latter it doesn't, so the $\s$-models don't seem to be isomorphic.  } \commentC{how is this possible if the complex structure is the same which fixes the torsion?} \comment{I think they represent flat space with two different complex structures, because the coordinate transformation $\tilde z^i = z^i/R^2$ is not holomorphic. }}{}
For $\a \neq 1$ the metric has  a curvature singularity in $R=0$ as can be seen from the Ricci-scalar 
$\calr = (D-1) (D-2) (1-\a^2) R^{-2 \a}$.
The coordinate transformation
\be 
R = (\a r)^{1\over \a}, %\qquad \a = 1-\d,
\ee
brings the metric in the canonical form (\ref{conecoords})
\be ds^2 = dr^2 + \a^2 r^2 d\O_{D-1}^2
\ee
with $d\O_{D-1}^2$ the round metric on the  sphere $S^{D-1}$.
The conformal Killing vector $\xi$, Reeb-like vector $\r$ and special conformal generator $K$ are given by
\be
\x = {1\over \a} \left( z^m \pa_m + \bar z^{\bar m} \pa_{\bar m}\right)\qquad 
\r =- {i\over \a} \left( z^m \pa_m - \bar z^{\bar m} \pa_{\bar m}\right)\qquad
K = {|z|^{2\a} \over 2\a^2}.\label{xirhotors}
\ee
The metric (\ref{metrtorsex})  possesses additional commuting real-holomorphic isometries which can be used to further refine the index. We can consider a separate rotation of the $m$-th complex plane  generated by
\be 
j_m = -i (z^m \pa_m - \bar z^{\bar m} \pa_{\bar m}).\label{jmdef}
\ee 
We will denote by  $\calj_m =- i L_{j_m}$  the corresponding symmetry  generator on the Hilbert space, acting as a spinorial Lie derivative.  One can check that these generators commute with $\calg_{\pm \half}$ and can be used to refine the index. Since the $u(1)_J$ generator is the combination \be J = {1 \over \a} \sum_{m=1}^{d_\CC} \calj_m, \label{JitoJm}\ee
we can refine the index by keeping track of $d-1$ additional charges. We choose those to be  $\calj_2, \ldots , \calj_d$ and will thus consider the refined index
\be 
\Omega_\pm [ q, x_2, \ldots, x_d]:= \tr (-1)^F e^{-\b \calh_\pm} q^{ J} \prod_{m=2}^d x_m^{\calj_m}.
\ee

To compute the index using localization techniques, we will again be interested in sigma models on a   target space where the conical singularity in the origin is resolved. Analogous to the $D=2$ examples we can replace the metric (\ref{metrtorsex}) by 
\be 
ds^2_{\rm reg} = f(R)^2  d z^m d\bar z^{\bar m}
\ee
with the function $f$  (as in the 2D examples) interpolating between 1 at the origin and $R^{\a-1}$ for $R \to \infty$ (see (\ref{bump}) for an explicit choice). We will treat the conformal and regularized models simultaneously,  omitting for simplicity  the notation $^{\rm reg}$ for the  regularized quantities. The   conformal model is simply recovered by setting $f=R^{\a-1}$. With respect to the vielbein  $e^{\underline{ m}} = f(R) dz^m$
one finds for the complex supercharges
\be 
\calq  = {1\over \sqrt{2}} f^{ F - {D \over 4}}\G^{ m} \pa_{  m} f^{ F + {D \over 4}}, \qquad \calq^\dagger  = -{1\over \sqrt{2}} f^{ -F - {3 D \over 4}}\G^{\bar m} \pa_{ \bar  m} f^{ -F + {3 D \over 4}}.
\ee
 Using  the explicit representation  (\ref{Gsexpl}) for the Dirac matrices, one checks that   for $D=2$ this reduces to our earlier result (\ref{calqregd2}). 
The  $u(1)$ generators $\calj_m $ are realized as
\be
\calj_m = -\left(z^m \pa_m - \bar z^{\bar m} \pa_{\bar m } + \half 1 \otimes \ldots \otimes \s_3 \otimes 1 \ldots 1\otimes 1\right) \qquad {\rm (no\ sum\ over\ }m),\label{Jmaction}
\ee
(where $\s_3$ appears in the $m$th position) and indeed commute with  $\calq_{}$.
We see that the definition (\ref{Fdef}) of $F$ implies that
\be 
F |\uparrow \uparrow \ldots \uparrow \rangle = {D\over 2} |\uparrow \uparrow \ldots \uparrow \rangle\qquad F |\downarrow \downarrow \ldots \downarrow \rangle = 0.
\ee
Using these results one finds that the chiral and anti-primary wavefunctions are of the form
\bea
| \chi^{ +}_{n_1, \ldots, n_{D/ 2}} \rangle&=& \left(\bar z^{ 1}\right)^{n_1} \ldots \left(\bar z^{D/ 2}\right)^{n_{D/ 2}}f^{- {D\over 4}} e^{- K^{\rm reg}} |\downarrow \downarrow \ldots \downarrow \rangle\\
| \chi^{-}_{n_1, \ldots, n_{D/ 2}} \rangle&=& \left( z^{ 1}\right)^{n_1} \ldots \left( z^{D / 2}\right) ^{n_{D/ 2}}f^{- {D\over 4}} e^{- K^{\rm reg}} |\uparrow \uparrow \ldots \uparrow \rangle.
\eea
Let us now discuss for which values of the integers $n_1, \ldots, n_{D/2}$ these states are actually normalizeable in the regularized and conformal models. The situation  for $D>2$ is quite different than that for $D=2$ discussed in Section \ref{Sec2D}. Indeed, in computing the norm of the BPS states we encounter integrals of the form (where $r_i \equiv |z_i|$)
\be 
|| | \chi^{{\rm reg}, \pm}_{n_1, \ldots, n_{D/ 2}} \rangle||^2 \sim \int dr_1\ldots dr_{D/2} r_1^{1+2 n_1}\ldots  r_{D/2}^{1+2 n_{D/2}} f(R)^{D\over 2} e^{-2 K_{\rm reg} (R)}.
\ee
For $D>2$, the strongest conditions on convergence come from considering the region where one of the $r_i$ goes to zero while the other $r_{j\neq 1}$ stay finite. Convergence of the integral in these regions  imposes that all the integers $n_i$ are positive, for the regularized as well as  the conformal models. Therefore, for $D>2$ the regularized model always correctly captures the BPS spectrum and the index, independent of the value of $\a$.  

%\change{Even though the relevant Dirac operator ceases to be essentially self-adjoint for sufficiently large $\a$ (namely for $\a > D-1$, assuming the torsion does not influence this value, the ambiguity in choosing a self-adjoint extension does not influence the spectrum of zero modes in this case, in contrast to what we found for $D=2$. \comment{Still confused about this. Another possibility  is that the torsion   alters  Chou's criterion so that the charge is essentially selfadjoint for all $\a$.}}{
This analysis suggests that  for $D>2$ the Dirac operator $\slashed{D}^{\rm tors}_{\cala^\pm}$ is essentially-selfadjoint over  the whole positive range of $\a$. While we will not attempt to prove this property in general, let us further justify this somewhat superprising conclusion. Indeed,  the naive application of Chou's criterion (\ref{Choucrit}) would predict that   the standard Dirac operator $\slashed{D}$ ceases to be essentially-selfadjoint for \be \a > D-1\label{Choubound}.\ee 
For larger values, negative states appear, i.e. zero modes which blow up at the singularity while remaining square integrable. The inclusion of torsion $C$  would somewhat modify this bound, in fact for D=4 would change it to $\alpha >2$, but would not eliminate it altogether.  We should however remember   that Chou's analysis is {\em local} near the singularity, as appropriate for conical singularities in {\em compact} manifolds, where convergence at infinity is not an issue. In our noncompact situation, we see from (\ref{calgsim}) that  chiral primary (resp. anti-chiral primary) states can be square-integrable for large $r$ only if they are built 
on the spin-down state  $ |\downarrow \downarrow \ldots \downarrow \rangle$ (resp. the spin-up state $|\uparrow \uparrow \ldots \uparrow \rangle$). We checked in D=4    that the   negative states  in Chou's analysis which appear beyond the bound (\ref{Choubound}) are in fact built  the mixed type spinors $|\downarrow \uparrow \rangle$ and $|\uparrow \downarrow \rangle$, and do not lead to normalizeable BPS states in our context. 
%The presence of torsion and background 
%, though we shall not attempt to prove this here. 
It would be interesting to have a more general understanding of the required modification of Chou's criterion (\ref{Choucrit}) %(see \cite{Chou1989}) 
to Dirac operators with torsionful connection and gauge bundle twist on noncompact cones.
%} 

From (\ref{Jmaction},\ref{JitoJm}) we find the quantum numbers of the BPS states under the various $u(1)$ generators:
\be 
\calj_m | \chi^{ \pm}_{n_1, \ldots, n_{d_\CC}} \rangle =\pm \left( n_m + \half \right)| \chi^\pm_{n_1, \ldots, n_{d_\CC}} \rangle\qquad J | \chi^{ \pm}_{n_1, \ldots, n_{d_\CC}} \rangle =\pm {1\over \a}\left( \sum_m n_m + {d_\CC\over 2}\right)| \chi^\pm_{n_1, \ldots, n_{d_\CC}} \rangle
\ee
and we find, for $D>2$, the unambiguous result for the index
\be 
\O^{\rm reg}_\pm[q, x_2,\ldots x_{d_\CC}] = \left[ {q^{1 \over 2 \a} \over 1-q^{1\over \a}} \prod_{I =2}^{d_\CC} {q^{1 \over 2 \a}x_I^\half \over 1-x_I q^{1\over \a}} \right]_\pm.\label{Omtorsex}
\ee

 Having computed  the indices using brute force, let us also illustrate how they can be efficiently obtained in the resolved models from a localization formula.  For this we take the fugacities to be pure phases and express the index  as
 \be 
 \O_\pm  [ e^{i \theta}, e^{ i\n^2}, \ldots ,e^{ i\n^d}] = \tr (-1)^F  e^{- \b \calh_\pm} g  (\theta, \n^I),
 \ee
 where $g (\theta, \n_I)$ is the group element
 \be 
 g (\theta, \m^I) = e^{i ( \theta J + \sum_{I =2}^{d_\CC} \n^I \calj_I)}.
 \ee
The index   localizes on the vanishing locus of the generator
\be 
j_{\theta, \m_I} =  \theta \r + \sum_{I =2}^{d_\CC}  \n^I j_I.
\ee
It's straightforward to see that there is a single isolated fixed point at $R=0$. 
%\comment{Assuming here that $f$ has no zeroes, might wonder what happens otherwise.} \commentP{I'd say that $f$ better not be zero, otherwise the regularization makes the metric degenerate.} %The infinitesimal  action on the tangent space at the fixed point, in the basis $\pa_1, \ldots , \pa_{d_\CC}$, is
%\be 
%\call_{j_{\theta, \m^I}}= {\rm diag} \left[i \left({  \theta   \over \a}, {  \theta   \over \a} + \m^2, \ldots ,  {   \theta   \over \a} + \m^{d_\CC} \right)\right]
%\ee
From (\ref{xirhotors}, \ref{jmdef}) we read off  the exponents of the various $u(1)$ actions at the fixed point $z^m =0$:
\bea
(l_\r)_a  &=& {1 \over \a}, \qquad a = 1 , \ldots , d_\CC\\
(l_I)_a &=& \d_{Ia}, \qquad I = 2 , \ldots , d_\CC.
\eea
Plugging these into the Atiyah-Bott fixed point formula (\ref{ABformisolated}) for the index  immediately leads to the result (\ref{Omtorsex}).

\section{K\"ahler targets and a type A/B relation }\label{Seckahler}

In this section we will consider the special class of type B sigma models where the target space is  K\"ahler and where the torsion tensor $C_{ABC}$ vanishes. We start with an example, and compute the superconformal index on the orbifold $\CC^2 /\ZZ_2$ using localization formulas on its resolution as the Eguchi-Hanson space (a similar computation for the conifold is discussed in Appendix \ref{Appcon}). 
In general, K\"ahler   manifolds also serve as target spaces for type A models which possess additional supersymmetries. This  leads to the question how the indices of the  type A and B  models defined on the same target space are related. In Section \ref{SecABrel} we derive a general relation between the type A and type B superconformal indices, and identify the type B  index on a Calabi-Yau cone with the Hilbert series of this singular space.

%\red{
Let us comment on the fact that, due to the absence of torsion, in these cases the criterion \eqref{Choucrit} applies to the study of possible ambiguities in the BPS spectrum near the conical singularity, at least if we restrict to vanishing background gauge field. On the K\"ahler quotient $\mathbb{C}^2/\mathbb{Z}_2$ (more generally $\mathbb{C}^D/\mathbb{Z}_N$), the base manifold is the quotient of a round sphere, so the eigenvalues of Dirac operator will be greater than the ones on the sphere. Thus we expect the index to be unambiguously reproduced by the smoothening procedure.
%}

\subsection{An example: $\mathbb{C}^2/\mathbb{Z}_2$ }

First we consider the example of the type B sigma model on the orbifold $\CC^2/\ZZ_2$. Choosing  coordinates $(w^1, w^2)$ on $\CC^2$, the identification acts  
as \be (w^1 , w^2 )\sim - (w^1 , w^2 ).\label{Z2id}\ee
The metric on the covering space $\CC^2$ is taken to be flat. A useful choice of coordinates is
\be
w^1 = R \cos{\theta \over 2}e^{-{i\over 2} (\psi + \f)},\qquad w^2 = R \sin{\theta \over 2}e^{-{i\over 2} (\psi - \f)},\label{EHws}
\ee
which, choosing  $R =r$, brings the metric in canonical form (\ref{conecoords})
\be 
ds^2 = dr^2 + {r^2\over 4} \left( d\theta^2 + \sin^2 \theta d\f^2+ (d\psi + \cos \theta d\f)^2\right)
 \ee
The $\ZZ_2$ identification (\ref{Z2id}) is reflected in the $\psi$-periodicity
\be 
\psi \sim \psi + 2 \p
\ee
being half of what it is on the three-sphere, reflecting the fact that the base of the cone is $S^3 /\ZZ_2$.

This target manifold satisfies the conditions for admitting  a superconformal  sigma model, where conformal Killing vector $\x$  and Reeb  vector $\r$ are
\bea 
\x &=& w^1 \pa_{w^1} + w^2 \pa_{w^2} + {\rm c.c.}=  r\pa_r \\ 
\r &=& -i( w^1 \pa_{w^1} + w^2 \pa_{w^2}) + {\rm c.c.} = 2 \pa_\psi.
\eea
Furthermore, there is an additional Killing vector 
\be 
j = \pa_\f = -{i\over 2} (w^1 \pa_{w^1} - w^2 \pa_{w^2}) + {\rm c.c.},
\ee
which satisfies the conditions (\ref{extrasymmconds}) to  define a second commuting $u(1)$ symmetry. We can also turn on a background gauge field while preserving  the conditions for conformal  invariance (\ref{N2conds},\ref{Rconds}). We take this to be an
$m$-monopole gauge potential\footnote{While  there is no a priori reason for $m$ to be quantized on the singular space $\CC^2/\ZZ_2$, only integer values of $m$ are compatible with the resolution discussed below.  The singularity will be replaced with a non-contractible 2-sphere, and flux quantization requires integer $m$.  } 
 \be 
 A = {m \over 2} (\pm 1 - \cos \theta)d\f.\label{AEH}
 \ee

A resolution of the singularity preserving the  structure required to define a regularized index replaces the orbifold singularity with an $S^2$. 
The resolution admits a Ricci-flat hyper-K\"ahler metric, namely the Eguchi-Hanson metric \cite{Eguchi:1978gw}
\be 
ds^2_{\rm reg} =  {dr^2 \over 1- \left({a \over r}\right)^4} + {r^2\over 4} \left( d\theta^2 + \sin^2 \theta d\f^2+ \left(  1- \left({a \over r}\right)^4\right)(d\psi + \cos \theta d\f)^2\right), \qquad r\geq a .\label{EHmetric}
\ee
Local complex coordinates $(w^1, w^2)$ can again be chosen to be of the form (\ref{EHws}), where $R$ is now related to $r$ as
\be 
R = (r^4 - a^4)^{1\over 4}.
\ee
From (\ref{SecABrel}) we see that the locus $r=a$  is a two-sphere where the $(w^1, w^2)$ coordinate system breaks down. A good system of coordinates \cite{Candelas:1987is} covers the manifold with two patches with coordinates $(z^1, z^2)$ and $(u^1, u^2)$ respectively:
\bea 
(z^1, z^2) &=& \left((w^1)^2, {w^2 \over w^1}\right ) \qquad {\rm for \ } \theta \neq \p\nonu
(u^1, u^2) &=& \left((w^2)^2, {w^1 \over w^2}\right ) \qquad {\rm for \ } \theta \neq 0\label{patchesEH}
\eea
On the overlap region, these  are related as
\be 
(z^1, z^2) = \left( u^1 (u^2)^2, {1\over u^2} \right).
\ee
These reflect the property that the Eguchi-Hanson space is the  total space of the bundle $T^* (S^2) \sim \calo(-2)$, with $z^2$ resp. $u^2$ local coordinates on the $S^2$ base manifold and   $z^1$ resp. $u^1$  being fiber coordinates. 

 The Killing vectors  $\r^{\rm reg} = 2\pa_\psi$ and $j^{\rm reg}=\pa_\f$ generate commuting real-holomorphic $u(1)$ symmetries. 
 For later convenience we list their component expressions  in both coordinate  patches (\ref{patchesEH}):
\bea 
\r^{\rm reg} &=& -2 i z^1\pa_{z^1} + + {\rm c.c.} = -2 i u^1\pa_{u^1} + {\rm c.c.} \\
j^{\rm reg} &=& - i\left(  z^1\pa_{z^1} -  z^2\pa_{z^2}\right)  + {\rm c.c.}  =  i\left(  u^1\pa_{u^1} -  u^2\pa_{u^2}\right)  + {\rm c.c.}\label{KVEH}
\eea
We take the background magnetic field on the Eguchi-Hanson space to be of the same form as (\ref{AEH}),
 \be 
 A^{\rm reg} = {m \over 2} (\pm 1 - \cos \theta)d\f.\label{AregEH}
 \ee
 We note that $j^{\rm reg}$ is associated through (\ref{extrasymmconds}) to  a nontrivial moment map which we take to be
\be 
\m_{j^{\rm reg}} = {m\over 2} \cos \theta.
\ee
The model on the Eguchi-Hanson space with  gauge  field (\ref{AregEH}) satisfies all the necessary conditions for  $N=2$ supersymmetry with two  additional central $u(1)$'s, and we can define   a refined index as %, and $ J^{\rm reg}, \calj^{\rm reg}$  which enter in the refined 
 \be 
 \O_\pm^{\rm reg}  [ q, x] = \tr (-1)^F  e^{- \b \calh_\pm}q^{ J^{\rm reg}} x^{\calj^{\rm reg}}.\label{OmregEH}
 \ee

\subsubsection{Unrefined index}
Let us first compute the  index without the  refinement by $\calj^{\rm reg}$ , setting $x=1$. It localizes on the zero locus of the  vector $\r^{\rm reg}$, which we see from (\ref{EHmetric}) is a two-sphere. Therefore we need the equivariant localization formula (\ref{ABformnonisolated}) for non-isolated fixed points, 
which reduces in this case to
\be 
\O [e^{i \theta}]= \int_{S^2} e^{c_1( F^{\rm reg}) } \left( 2 \sinh (  c_1 (N) +  l_{\r^{\rm reg}}  \theta)/2 \right)^{-1}\label{localS2unref}
\ee
Here, $l_{\r^{\rm reg}}$ is the exponent of the $\r^{\rm reg}$-action on the normal bundle  %(i.e. on the basis vector $\pa_{z^2}$ or $\pa_{u^2}$), and   $c_1 (F)$ 
and 
$c_1 (N)$ is the first Chern class of the latter. More precisely we have
\be 
l_{\r^{\rm reg}} = 2  , \qquad  \int_{S^2}c_1( F^{\rm reg}) =-m, \qquad \int_{S^2} c_1 (N) =-2,
\ee
where the last equality is due to the fact that, as already mentioned, the manifold is the total space of the bundle $\calo (-2)$ over $S^2$. Picking out the contributing  terms in (\ref{localS2unref}) leads to 
\be
\O[q]^{\rm reg} =  {q(1+m + (1-m) q^2)\over (1-q^2)^2 } .\label{EHindunref}
\ee

\subsubsection{Refined index}
We now compute the refined index $\O_\pm^{\rm reg} [q,x]$ defined in (\ref{OmregEH}).
Taking $q = e^{i \theta}, x  = e^{i \n }$, the index localizes on the vanishing locus of the generator
\be 
j_{\theta, \n} =   \theta \r^{\rm reg} +  \n j^{\rm reg}.
\ee
When $\n \neq 0$, it is straightforward to see that there are two isolated fixed points at  
$z^m =0$ and at  $u^m =0$, i.e.  at the north and south poles of the two-sphere at $r=a$. From (\ref{KVEH})  we read off the exponents of the $u(1)$ action and values of the moment map at these fixed points:
%\bea 
%\call_{j_{\theta, \n}}&=& {\rm diag} \left[i \left( 2 i \theta + i\n, -i \n \right)\right] \qquad {\rm at\ } z^m =0\\
%&=& {\rm diag} \left[i \left( 2 i \theta - i\n , i\n \right)\right] \qquad {\rm at\ } u^m =0
%\eea
\begin{align}
{\rm at\ } z^m =&0: & l_{\r^{\rm reg}} =& (2,0) , & l_{j^{\rm reg}} =&(1,-1), &\m_{j^{\rm reg}} =& {m\over 2} \\
{\rm at\ } u^m =&0: & l_{\r^{\rm reg}} =& (2,0) , & l_{j^{\rm reg}} =&(-1,1), &\m_{j^{\rm reg}} =& -{ m\over 2} 
\end{align}
Substituting this data in the Atiyah-Bott formula  (\ref{ABformisolated}) one obtains
 \be
\O^{\rm reg} [ q,x] = {q  x^{-{m\over 2}}\over 1-x }\left( {1\over  1- q^2 x^{-1}} -  {x^{m+1}\over  1- q^2 x }\right)
\ee 
One checks that, in the limit $x \to 1$, this reduces to the unrefined result (\ref{EHindunref}).
Turning off the background field, $m=0$, leads to   
\be 
\O^{\rm reg} [ q,x]_{| m=0} = { q (1+q^2)\over \left( 1- q^2 x^{-1}\right)(1- q^2 x)  },\label{refindEHm0}
\ee
which, as we will see in the next subsection, can be compared  with the result of \cite{Dorey:2019kaf} for the type A index.

The above analysis can be generalized in a straightforward manner to singular quiver varieties  whose resolution is the cotangent bundle of $\CC P^{N}$ (see Section 5 in \cite{Dorey:2019kaf}). As an  example which is neither an orbifold nor hyperk\"ahler, we discuss in Appendix \ref{Appcon}   the superconformal index on the conifold, which we smoothen out using its small resolution. In target spaces  that can be  described as toric varieties, such as the Eguchi-Hanson space and the conifold,    the exponents of the $u(1)$ actions at the fixed points needed to compute the refined index can be simply read off from the toric data.% We will return to this description in the next section.

 \subsection{Relation between type A and type B indices}\label{SecABrel}
In the  previous subsection and Appendix \ref{Appcon} we have computed the type $2$B superconformal index of sigma models with  K\"ahler targets and with vanishing torsion tensor $C$.  Such manifolds can also serve as target spaces of type A sigma models \cite{Dorey:2019kaf}. As already mentioned, type A theories  
are realized on multiplets with 2 bosons and 4 fermions and posses additional supercharges constructed from the extra fermions. Type A models on K\"ahler manifolds  possess at least 4 supercharges, and in the superconformal case the full algebra includes $u(1,1|2)$ \cite{Dorey:2019kaf}. 

In this subsection we work out in more detail how the type A and type B models on a K\"ahler manifold are related. We will see that the type $2$B Hilbert space is embedded as a subspace of the type $4$A Hilbert space at fixed value for one of the $R$-charges. The $u(1,1|1)\oplus u(1)$ symmetry of the type B model is the isotropy subalgebra of  this charge in $u(1,1|2)$.
 This embedding will allow us to express the type B superconformal index on K\"ahler manifolds as a specific limit of the type A index.

The relation between the two types of models is most transparent in the representation of their Hilbert spaces as spaces of differential forms. Let us first recall this representation for the type B models. 
Here, we can take the Hilbert space to be the space of $(0 , \bullet )$ forms with the usual inner product.
The Clifford algebra elements are represented as
\be 
\G^{\bar m} =\sqrt{2} e^{\bar m} \wedge, \qquad \G^m = \sqrt{2} g^{m \bar n} {\d \over \d e^{\bar n}}
\ee
One checks (see  \cite{Raeymaekers:2024usy} for details) that the superconformal charges and $u(1)_J$ generator are represented as
\bea 
\left(\calg_{\pm \half}\right)^\dagger &=& - e^{\pm K} (\sqrt{G})^{-{1\over 4}}\left( \bar \pa - i A_{(0,1)}\wedge \right)  (\sqrt{G})^{{1\over 4}}  e^{\mp K} \label{calgdiff} \\
J &=& - i\call_\r - \r^A A_A - {d_\CC \over 2},
\eea
where $\call$ denotes the Lie derivative acting on differential forms. 
Furthermore, the fermion number operator $F$ acts as
\be 
F = d_\CC -q
\ee
on $(0,q) $ forms.
For  vanishing background gauge potential $A_A$, which we will focus on for the rest of this subsection, the superconformal index counts Dolbeault cohomology classes which are normalizeable with respect to a nonstandard measure, on which $J$ acts (up to a constant) as the Lie derivative.

For the type A model on the same target space, the Hilbert space consists of general $(p,q)$ forms  as the additional fermionic generators mix   forms with different values of $p$. The type A  superconformal  index is based on a supercharge which \cite{Dorey:2018klg} again  acts  as $\bar \pa $ up to a similarity transformation as in (\ref{calgdiff}). 
The authors of \cite{Dorey:2019kaf} define the refined superconformal type A index as
\be 
\calz [q, y, x_I] = \tr_{\calh_A} (-1)^{F_A} e^{-\b H_A} q^{D^I}y^{J_3 + R^I } \prod_I x_I^{\calj_I},
\ee
where $\calh_A$ is the type A Hilbert space, the `Hamiltonian' $H_A$ is given by
\be 
H_A = L_0 + J_ 3 - {D^I\over 2},
\ee 
and the various charges act on $(p,q)$-forms as
\begin{align}
F_A &= p+ q - d_\CC& J_3 &=\half( p + q - {d_\CC }  )\\
D^I &=  i\call_\r  & R^I &=\half(p-q).
\end{align}
Restricting to the type B subspace consisting of forms with $p=0$, we see that $H_A$ reduces to our $\calh_-/ 2$, and that
the type B index $\O_-$  is related to  the coefficient of $y^{- d_\CC/2}$ in $\calz$. More precisely the type A and B indices are related as
\be 
\O_- [q^{-1},x_I] =  \lim_{y\to 0} (y q)^{d_\CC\over 2}  \calz [q, y, x_I].\label{ABrel}
\ee
We can check this relation for the examples in the last subsection and Appendix \ref{Appcon}: indeed, the type B index on $\CC^2 / \ZZ_2$ (\ref{refindEHm0}) agrees with eq. (5.27) in \cite{Dorey:2019kaf}, while the index on the conifold (\ref{refindconm0}) agrees with eq. (5.6) in that work.  

Using the relation (\ref{ABrel}) and the results of  \cite{Dorey:2019kaf} allows us, in certain cases, to characterize the type B index directly in terms of intrinsic geometric properties of the singular space. Indeed, for target spaces which are Calabi-Yau cones, it was shown in \cite{Dorey:2019kaf} that the limit appearing on the RHS of (\ref{ABrel}) %precisely 
coincides with the Hilbert series of the singular space, up to a factor of $q^{d_{\mathbb C}/2}$. The type B index  in this case counts holomorphic functions on the singular cone and is  manifestly  independent of the way we resolved the target space.

 \section{Discussion}
 In this work we have proposed a method for computing the superconformal index of type B models on singular target spaces. It involves  working on a regularized   target space obeying suitable differential-geometric properties and is similar in spirit to the work on the type A index  \cite{Dorey:2018klg, Barns-Graham:2018xdd, Dorey:2019kaf} but relies on  techniques from differential rather than  algebraic geometry.   We also pointed out a potential ambiguity in the index when the relevant supercharge is not essentially-selfadjoint. 
 In this context, it would be interesting to have a more general understanding of the required modification of Chou's  \cite{chou1985dirac} criterion (\ref{Choucrit}) 
to Dirac operators with torsionful connection and gauge bundle twist on noncompact cones.

 One of the primary motivations for this research was to facilitate the calculation of superconformal indices for quiver quantum mechanics in the Coulomb branch \cite{Denef:2002ru}, particularly within an AdS$_2$ scaling limit. These indices can be expressed as $N=4$B sigma models where a $U(1)$ symmetry is gauged \cite{Mirfendereski:2022omg} and which possess $D(2,1;0)$ superconformal invariance \cite{Anninos:2013nra,Mirfendereski:2020rrk}.  We will study these quiver theories in a future publication \cite{wip-quivers}. The exploration of these models is anticipated to provide insights into to explicit description of black hole microstates as D-brane bound states and the stringy origins of AdS$_2$/CFT$_1$ duality.

\acknowledgments

It is a pleasure to thank Nick Dorey, Chris Pope, and Andy Zhao for useful discussions. We are especially grateful to Dieter Van den Bleeken for initial collaboration on this project. The work of C\c{S} was supported by the European Union's Horizon Europe programme under grant agreement No. 101109743, project Quivers. He is also grateful to the DAMTP, University of Cambridge for their hospitality. P.R. was supported by the Grant Agency of the Czech Republic under the grant
EXPRO 20-25775X. The research of J.R.  was supported by European Structural and Investment Funds and the Czech Ministry of Education, Youth and Sports (Project FORTE CZ.02.01.01/00/22\_008/0004632).

\begin{appendix}

\section{Self-adjointness in 2D models} \label{app:selfadjoint_supercharge}

\subsection{Supercharge on 2D cones}

%\commentP{ Refs, the analysis follows reasonings in Falomir-Pisani, Kay-Studer, Reed-Simons,...} 
Let us consider the general 2D models studied in Section \ref{Sec2D}. In radial coordinates $(r,\theta)$ where the angle coordinate is subject to identification $\theta\sim\theta+2\pi\alpha$, the metric takes the form $ds^{2}=r^{2}+r^{2}d\theta^{2}$. Consider the hermitian combinations of supercharges \eqref{gpmvan} expressed in radial coordinates,\footnote{The other hermitian combination $G_2^{s}=i(\mathcal{G}_{s/2}-\mathcal{G}_{s/2}^{\dagger})$, which was identified in \eqref{eq:Dirac_op_wtorsion} with a twisted Dirac operator, is related to $G_1^{s}$ by the unitary transformation $\tilde{\mathcal{Q}}_{s}=e^{i\pi\sigma^{3}/4}G_1^{s}e^{-i\pi\sigma^{3}/4}$. So we need only to concentrate on one of the two.}
%\commentP{Here assumed $\sigma_{\pm}=\frac{1}{2}(\sigma^{1}\pm i\sigma^{2})=|\pm\rangle\langle\mp| $. Adjust notation for spin part. Also symbol $calq$ was already used, to fix}
\begin{equation}
G_1^{s}:=\mathcal{G}_{s/2}+\mathcal{G}_{s/2}^{\dagger}=\frac{1}{\sqrt{2}}\sum_{\sigma\in\{\pm\}}e^{i\sigma\theta/\alpha}|-\sigma\rangle\langle\sigma|\left(-\sigma\partial_{r}+\frac{1}{ir}\partial_{\theta}+\sigma\frac{1-\alpha}{2\alpha}\frac{1}{r}+sr\right),
\end{equation}
where $s=\pm$.
This formal charge acts on a dense subspace of the spinorial Hilbert space $L^{2}(\mathbb{R}^{2}\setminus\{0\},rdr\frac{d\theta}{2\pi\alpha})\otimes\mathbb{C}^{2}$, yet to specify precisely.
%\commentP{ topologically the smooth part of the domain is a cylinder, but the angular coordinate has period $2\pi\alpha$. Maybe we should make this clear by replacing $\mathbb{R}^{2}\setminus\{0\}$ with some notation for ``planar wedge''? This is only a notation issue}. Notice that, since the point $r=0$ is singular in the geometry, we had to remove it from the domain of the wavefunctions in order to study their analytical properties. On the smooth domain $\mathbb{R}^{2}\setminus\{0\}$, the radial coordinates $(r,\theta)$ are globally defined. 
By separation of variables, we can expand wavefunctions and the action of the supercharges as 
\begin{align}
\Phi(r,\theta) & =\sum_{l\in\mathbb{Z}}\sum_{\sigma\in\{\pm\}}\phi_{l}^{\sigma}(r)e^{il\theta/\alpha}|\sigma\rangle\\
G_1^{s}\Phi(r,\theta) & =\sum_{l,\sigma}\frac{-\sigma}{\sqrt{2}}\left(\partial_{r}-\frac{c_{l\sigma}(\alpha)}{r}-s\sigma r\right)\phi_{l}^{\sigma}(r)e^{i(l+\sigma)\theta/\alpha}|-\sigma\rangle\qquad\Big(c_{l\sigma}(\alpha):=\frac{2l\sigma+1-\alpha}{2\alpha}\Big).
\end{align}

Because the domain $\mathbb{R}\setminus\{0\}$ is open, one should impose appropriate boundary conditions on the behavior of the wavefunctions at $r\to0$ (requiring square-integrability in the radial measure $rdr$ fixes uniquely the behavior at the other ``boundary'' $r\to\infty$). The key idea is that we want to realize the supersymmetry algebra %$\{G_1^{s},\mathcal{H}_{s}\}$ 
on the Hilbert space as an algebra of physical, \emph{self-adjoint} operators. This leads to the precise behavior of the wavefunctions near the conical singularity, and consequently to a precise account of the BPS spectrum. 

Let us notice that if $G_1^{s}$ is self-adjoint on some dense domain $\mathcal{D}\subset L^{2}(\mathbb{R}^{2}\setminus\{0\},rdr\frac{d\theta}{2\pi\alpha})\otimes\mathbb{C}^{2}$, then the Hamiltonian $\mathcal{H}_{s}=(G_1^{s})^{2}$ is also self-adjoint on the domain $\mathcal{D}(\mathcal{H}_{s})=\{\Psi\in\mathcal{D}|G_1^{s}\Psi\in\mathcal{D}\}$ but the contrary is not true. Infact, in general there can be domains of self-adjointness for the Hamiltonian which are not well-defined for the supercharge. %\commentP{comment on this at the end by quickly study the Hamiltonian operator?}. 
Since our aim is to preserve supersymmetry at the quantum level, we focus on  the supercharge and not on the Hamiltonian.

To start off, a standard choice for dense domain of $G_1^{s}$ is $\mathcal{V}=C_{o}^{\infty}(\mathbb{R}^{2}\setminus\{0\})\otimes\mathbb{C}^{2}$, the space of smooth wavefunctions compactly supported away from the origin. Since any $\Psi\in\mathcal{V}$ satisfies $\Psi(r,\theta)\xrightarrow{r\to0}0$, it is easy to show that $G_1^{s}$ is symmetric on this domain with respect to the $L^{2}$ measure $rdr$, due to vanishing of boundary terms after partial integration. %Below we apply the general strategy presented in the last section to the study of self-adjoint extensions of the densely-defined, symmetric operator $G_1^{s}$ and then analyze its BPS spectrum. Let us collect the final results here.
Starting from the symmetric operator $G_1^{s}$ on $\cal V$ and applying the general strategy presented in \ref{app:basic_notions_selfadj}, we show the following relevant facts. Detailed arguments are provided in the next section. 

\begin{enumerate}
\item When the tip-opening parameter $\alpha$ is in the range $(0,1]$, the supercharge $G_1^{s}$ is essentially self-adjoint. The modes $\phi_{l}^{\sigma}(r)$ of any wavefunction $\Phi\in\mathcal{D}$ in its domain of self-adjointness are absolutely continuous, square-integrable functions of the radial coordinate $r\in\mathbb{R}_{>0}$ with respect to the measure $rdr$, respecting the boundary behavior 
%{[}\textbf{P: is the distinction between $o(1)$ and $O(1)$ clear?}{]}
\begin{align}
\alpha\in(0,1): &\  \phi_{l}^{\sigma}(r)\approx o(1)\qquad\forall l,\sigma\\
\alpha=1: & \begin{cases}
\phi_{0}^{\sigma}(r)\approx O(1) & \forall\sigma\\
\phi_{l}^{\sigma}(r)\approx o(1) & \forall l\neq0,\sigma
\end{cases}
\end{align}
In particular, wavefunctions in the domain of $G_1^{s}$ are regular at the origin.

\item When the tip-opening parameter $\alpha$ is in the range $(1,\infty)$, the supercharge $G_1^{s}$ is not essentially self-adjoint, instead it has many self-adjoint extensions. In all of these, the modes $\phi_{l}^{\sigma}(r)$ of any wavefunction $\Phi\in\mathcal{D}_{\mathcal{U}}$ are absolutely continuous, square-integrable functions of the radial coordinate $r\in\mathbb{R}_{>0}$ with respect to the measure $rdr$. The regular boundary conditions
\begin{align}
\phi_{l}^{\sigma}(r) & \approx o(1)\qquad\text{for }l\sigma\geq\frac{\alpha-1}{2}\text{ and }l\sigma<-\frac{\alpha+1}{2},\\
\phi_{l}^{\sigma}(r) & \approx O(1)\qquad\text{for }l\sigma=\frac{\alpha-1}{2}\ \big(\text{only when }\alpha\in\mathbb{N}_{\mathrm{odd}}\big),
\end{align}
are required in every domain of self-adjointness. Instead, modes in the range $-\frac{\alpha+1}{2}<l\sigma<\frac{\alpha-1}{2}$ can generically diverge at the origin as
\begin{equation}
\phi_{l}^{\sigma}(r)=\sum_{-\frac{\alpha+1}{2}<\overline{l\sigma}<\frac{\alpha-1}{2}}a_{\overline{l\sigma}}^{l\sigma}r^{c_{\overline{l\sigma}}(\alpha)}+o(1)\qquad\text{for }-\frac{\alpha+1}{2}<l\sigma<\frac{\alpha-1}{2},
\end{equation}
for some choice of coefficients $a_{\overline{l\sigma}}^{l\sigma}$. Different choices of self-adjoint extensions corresponds to choices of $a_{\overline{l\sigma}}^{l\sigma}$, up to an overall normalization.

\item In the conical defect range $\alpha\in(0,1]$, the BPS spectrum consists of the square-integrable states
\begin{equation}
\chi_{l}^{(s)}(r,\theta)\propto r^{c_{l(-s)}(\alpha)}e^{il\theta/\alpha}e^{-r^{2}/2}|-s\rangle\propto\begin{cases}
z^{l}|z|^{\frac{1-\alpha}{2}}e^{-\frac{|z|^{2\alpha}}{2\alpha^{2}}}|\uparrow\rangle & s=-,l\geq0\\
\bar{z}^{|l|}|z|^{\frac{1-\alpha}{2}}e^{-\frac{|z|^{2\alpha}}{2\alpha^{2}}}|\downarrow\rangle & s=+,l\leq0
\end{cases}
\end{equation}
where we reintroduced the complex coordinates $z=(\alpha re^{i\theta})^{1/\alpha}$ to recognize chiral and antichiral states \eqref{cp2d},\eqref{acp2d}.

\item In the range $\alpha>1$, every self-adjoint extension of $G_1^{s}$ has the BPS states
\begin{equation}
\chi_{l}^{(s)}(r,\theta)\propto r^{c_{l(-s)}(\alpha)}e^{il\theta/\alpha}e^{-r^{2}/2}|-s\rangle\qquad\forall l:sl\leq\frac{1-\alpha}{2}.
\end{equation}
The states $\chi_{l}^{(s)}$ for $l$ in the range $\frac{1-\alpha}{2}<sl<\frac{1+\alpha}{2}$ are divergent but square-integrable. Different choices of self-adjoint extensions correspond to choosing which linear combinations of the diverging states to include in the BPS spectrum.

\end{enumerate}

\subsection{Some proofs}

Below we review the steps to reach the claims made in the last section in some detail. Although the computations are adapted to our setup, the strategy follows standard ideas, see e.g.\ \cite{Falomir_2005,Kay:1991}.

\paragraph{The adjoint operator $(G_1^{s})^{*}$}

Here we study the domain $\mathcal{V}^{*}$ of the adjoint supercharge. Any $\Psi\in\mathcal{V}^{*}$ satisfies $\langle\Psi,G_1^{s}\Phi\rangle=\langle\Omega^{\Psi},\Phi\rangle$ for any $\Phi\in\mathcal{V}$ and some $\Omega^{\Psi}$ in the Hilbert space. Since $\Phi\in\mathcal{V}$ kills the boundary terms arising from partial integration, we see that the definition of adjoint operator implies
\begin{equation}
-\sigma\sqrt{2}(\Omega^{\Psi})_{l}^{\sigma}(r)=\partial_{r}\psi_{l}^{\sigma}(r)-\left(\frac{c_{l\sigma}(\alpha)}{r}+s\sigma r\right)\psi_{l}^{\sigma}(r)
\end{equation}
in a distributional sense. Since $\left(\frac{c_{l\sigma}(\alpha)}{r}+s\sigma r\right)\psi_{l}^{\sigma}(r)$ is locally $L^{2}$, this shows that all $\partial_{r}\psi_{l}^{\sigma}$ is locally $L^{2}$, hence $\psi_{l}^{\sigma}$ is an absolutely continuous (in particular continuous, regular) function on $\mathbb{R}_{>0}$. Therefore, the domain of the adjoint supercharge consists of those square-integrable wave functions $\Psi$ whose modes $\psi_{l}^{\sigma}(r)$ are absolutely continuous and such that $(\partial_{r}-\frac{c_{l\sigma}(\alpha)}{r}-s\sigma r)\psi_{l}^{\sigma}(r)$ is also square-integrable:
\begin{equation}
\mathcal{V}^{*}=\bigoplus_{l,\sigma}\left\{ \psi_{l}^{\sigma}\in L^{2}(\mathbb{R}_{>0},rdr)\cap AC(\mathbb{R}_{>0})\Big|\ (G_1^{s}\psi_{l}^{\sigma})\in L^{2}(\mathbb{R}_{>0},rdr)\right\} .
\end{equation}
We conclude that the action of the adjoint operator $(G_1^{s})^{*}$ is through the same differential operator as $G_1^{s}$, only interpreted in a distributional sense.

Let us examin what the above statement means practically in terms of the behavior of $\Psi\in\mathcal{V}^{*}$ near $r\to0$. For the modes $\psi_{l}^{\sigma}$ to be square-integrable near the origin in the measure $rdr$, it must be that $\psi_{l}^{\sigma}(r)\approx o(r^{-1})$ for $r\to0$. On the other hand, for $(G_1^{s}\Psi)_{l}^{\sigma}$ to be square-integrable near the origin it must be that $\psi_{l}^{\sigma}(r)\approx\#r^{c_{l\sigma}(\alpha)}+o(1)$. Enforcing both conditions at the same time, we find that for conical defects the behavior of the modes near the origin is
\begin{equation}\begin{aligned} \label{eq:bry_cond_adjoint_supercharge}
\alpha\in(0,1) & :\psi_{l}^{\sigma}(r)\approx o(1)\qquad\forall l,\sigma,\\
\alpha=1 & :\psi_{0}^{\sigma}(r)\approx O(1),\qquad\psi_{l\neq0}^{\sigma}(r)\approx o(1).
\end{aligned}\end{equation}
For conical surpluses the situation is more involved: every mode $\psi_{l}^{\sigma}$ with $-\frac{\alpha+1}{2}<l\sigma\leq\frac{\alpha-1}{2}$ (this range is nontrivial only when $\alpha\geq1$) can behave as 
\begin{equation}
\psi_{l}^{\sigma}(r)\approx\#r^{c_{l\sigma}(\alpha)}+o(1),
\end{equation}
since $-1<c_{l\sigma}(\alpha)\leq0$. All other modes need to decay as $o(1)$.

Already at this point we can appreciate that, for $\alpha\in(0,1]$ in the range of conical defects, the supercharge is essentially self-adjoint. This is because for the boundary conditions \eqref{eq:bry_cond_adjoint_supercharge} the adjoint charge $(G_1^{s})^{*}$ is symmetric and thus the unique self-adjoint extension of $G_1^{s}$. To see this, we should take two states $\Psi,\Phi\in\mathcal{V}^{*}$ and try to show $\langle\Psi,(G_1^{s})^{*}\Phi\rangle\stackrel{?}{=}\langle (G_1^{s})^{*} \Psi,\Phi\rangle$, which is obviously true up the boundary term
\begin{equation}
\sum_{l,\sigma}\frac{\sigma}{\sqrt{2}}\lim_{r\to0}\left[r\overline{\psi_{l+\sigma}^{-\sigma}(r)}\phi_{l}^{\sigma}(r)\right]
\end{equation}
popping up after partial integration (the term at $r\to\infty$ must give zero because wavefunctions are $L^{2}$). Using $c_{(l+\sigma)(-\sigma)}(\alpha)=-c_{l\sigma}(\alpha)-1$, it is easy to see that this is nonzero (and finite) whenever at least one of the modes of $\Psi$ and $\Phi$ diverges, while it is zero otherwise. So only the boundary conditions associated to the domain $\mathcal{V}^{*}$ for $\alpha\in(0,1]$ ensure $\langle\Psi, (G_1^{s})^{*}\Phi\rangle=\langle (G_1^{s})^{*}\Psi,\Phi\rangle$, making the supercharge symmetric.

\paragraph{The closure operator}

For later reference, let us work out the domain of the closure $\overline{G_1^{s}}=(G_1^{s})^{**}$. First, in the range $\alpha\in(0,1]$ where $G_1^{s}$ is essentially self-adjoint the closure coincides with the adjoint, which is infact self-adjoint. So we are left to consider the surplusses $\alpha>1$.

We can work out the closure operator as the double adjoint of the supercharge. Using this characterization, elements $\Psi\in\overline{\mathcal{V}}$ are defined so that $\langle\Psi,(G_1^{s})^{*}(\cdot)\rangle$ is a continuous linear functional on $\mathcal{V}^{*}$. Since $\overline{\mathcal{V}}\subset\mathcal{V}^{*}$, every $\Psi\in\overline{\mathcal{V}}$ is also in the domain of the adjoint. As such, its modes $\psi_{l}^{\sigma}(r)$ must be absolutely continuous in the radial coordinate $r\in\mathbb{R}_{>0}$ and square-integrable in the measure $rdr$, such that also $(G_1^{s}\Psi)_{l}^{\sigma}(r)$ are square-integrable. Thus integration by parts is justified, and for every $\Phi\in\mathcal{V}^{*}$ we can write
\begin{equation}
\langle\Psi,G_1^{s}\Phi\rangle=\langle G_1^{s}\Psi,\Phi\rangle+\sum_{\sigma,l}\frac{\sigma}{\sqrt{2}}\lim_{r\to0}\left[r\overline{\psi_{l+\sigma}^{-\sigma}(r)}\phi_{l}^{\sigma}(r)\right],
\end{equation}
where as usual the boundary term pops up after partial integration, and only the $r\to0$ term can be nontrivial. For $\langle\Psi,G_1^{s}(\cdot)\rangle$ to be a continuous functional, the boundary term needs to vanish.\footnote{If $\Phi_{n}\xrightarrow{n\to\infty}\Phi$ is a converging sequence in the Hilbert space, $\lim_{n\to\infty}\lim_{r\to0}\left[r\overline{\psi_{l+\sigma}^{-\sigma}(r)}(\phi_{l}^{\sigma})_{n}(r)\right]\neq\lim_{r\to0}\left[r\overline{\psi_{l+\sigma}^{-\sigma}(r)}\phi_{l}^{\sigma}(r)\right]$ because to exchange the limits we would need the convergence $(\phi_{l}^{\sigma})_{n}(r)\xrightarrow{n\to\infty}\phi_{l}^{\sigma}(r)$ to be \emph{uniform}, while we only have $L^{2}$-convergence. So the boundary term is not a continuous functional.} For those modes of $\Phi$ such that $-1<c_{l\sigma}(\alpha)<0$, which are allowed to diverge like $\phi_{l}^{\sigma}(r)\approx\#r^{c_{l\sigma}(\alpha)}+o(1)$, killing the boundary term requires $\psi_{l+\sigma}^{-\sigma}(r)\approx o(r^{-c_{l\sigma}(\alpha)-1})=o(r^{c_{(l+\sigma)(-\sigma)}(\alpha)})$. Since also $\phi_{l+\sigma}^{-\sigma}$ diverges as $r^{-c_{l\sigma}(\alpha)-1}$, dually $\psi_{l}^{\sigma}(r)\approx o(r^{c_{l\sigma}(\alpha)})$. For all these modes, square integrability of $(G_1^{s}\psi_{l}^{\sigma})$ actually forces $\psi_{l}^{\sigma},\psi_{l+\sigma}^{-\sigma}(r)\approx o(1)$. When $\alpha$ is an odd integer the modes with $l\sigma=\frac{\alpha-1}{2}$, which implies $c_{l\sigma}(\alpha)=0$, must be treated separately: the mode $\phi_{l}^{\sigma}(r)\approx O(1)$ is regular but finite at the origin, while $\phi_{l+\sigma}^{-\sigma}(r)\approx o(1)$ needs to decay. Consequently, to kill the boundary term and preserve square-integrability $\psi_{l+\sigma}^{-\sigma}(r)\approx o(1)$ and $\psi_{l}^{\sigma}(r)\approx O(1)$.

Summarizing, for any conical parameter $\alpha\in(0,\infty)$, the modes $\psi_{l}^{\sigma}(r)$ of a wavefunction  $\Psi\in\overline{\mathcal{V}}$ are square-integrable in the measure $rdr$, absolutely continuous in $r\in\mathbb{R}_{>0}$ and satisfy the boundary conditions
\begin{align}
\alpha\notin2\mathbb{Z}_{\geq0}+1 & :\psi_{l}^{\sigma}(r)\approx o(1),\\
\alpha=2n+1\text{ for }n\in\mathbb{Z}_{\geq0} & :\begin{cases}
\psi_{l}^{\sigma}(r)\approx O(1) & l\sigma=n\\
\psi_{l}^{\sigma}(r)\approx o(1) & \text{otherwise}.
\end{cases}
\end{align}
In particular any $\Psi(r,\theta)\in\overline{\mathcal{V}}$ is regular at $r\to0$.

\paragraph{Deficiency subspaces}

We look at the deficiency subspaces $\mathcal{K}_{\pm}=\ker((G_1^{s})^{*}\mp i)$, so the formal eigenvalue equation $G_1^{s}\Phi=\lambda\Phi$ with $\lambda=\pm i$ (we drop the $^{*}$-sign since we know that the adjoint acts as the same differential operator as $G_1^{s}$). Since $(G_1^{s})^{2}=\mathcal{H}_{s}$, any eigenstate of $G_1^{s}$ must also solve $\mathcal{H}_{s}\Phi=\lambda^{2}\Phi$ with energy $\lambda^{2}=-1$. On the other way, an eigenstate of the Hamiltonian generically is not in the domain of the supercharge. Only when this is true, the linear combination $(\lambda\Phi+G_1^{s}\Phi)$ is an eigenstate of $G_1^{s}$ with eigenvalue $\lambda$.

Because the Hamiltonian acts diagonally on the mode decomposition of the wavefunctions, it is easier to solve the latter eigenvalue equation and then analyze which of its solutions can give rise to eigenstates of the supercharge. The action of $\mathcal{H}_{s}$ decomposes as
\begin{equation}\begin{aligned}
\mathcal{H}_{s}\Phi(r,\theta) &= \sum_{l,\sigma}\frac{-1}{2}\left[\partial_{r}^{2}+\frac{1}{r}\partial_{r}-\frac{c_{l\sigma}(\alpha)^{2}}{r^{2}}-r^{2}-2sR_{l\sigma}(\alpha)\right]\phi_{l}^{\sigma}(r)e^{il\theta/\alpha}|\sigma\rangle \\
& \text{where } R_{l\sigma}(\alpha):=\sigma(1+c_{l\sigma}(\alpha)).
\end{aligned}
\end{equation}
Technically it is convenient to make the substitution $\phi_{l}^{\sigma}(r)=e^{-r^{2}/2}r^{|c_{l\sigma}(\alpha)|}F_{l}^{\sigma}(r^{2})$, so that the energy eigenvalue equation $\mathcal{H}_{s}\Phi=E\Phi$ reduces to \emph{Kummer's hypergeometric equation}
\begin{equation} \label{eq:Kummer_equation}
0=\left[x\frac{d^{2}}{dx^{2}}+\left(b_{l\sigma}(\alpha)-x\right)\frac{d}{dx}-a_{l\sigma}^{s}(\alpha;E)\right]F_{l}^{\sigma}(x).
\end{equation}
where $x\equiv r^{2}$, $b_{l\sigma}(\alpha)=1+|c_{l\sigma}(\alpha)|$ and $a_{l\sigma}^{s}(\alpha;E)=\frac{1}{2}\left(1+|c_{l\sigma}(\alpha)|+s\sigma(c_{l\sigma}(\alpha)+1)-E\right)$ are the relevant parameters. As a formal ODE of second order, this has two linearly independent local solutions, but depending on the parameters $a$ and $b$ some of these solutions may fail to be $L^{2}$ and have to be discarded. An accurate study of the solutions of Kummer's equation can be found in \cite{Mathews_2022Kummer_eq, NIST:DLMF}.

For the eigenvalue $E=\lambda^{2}=-1$, the parameter $\alpha_{l\sigma}^{s}(\alpha;\lambda^{2})$ can never be a negative integer. Therefore one finds that all locally square-integrable solutions are of the sort
\begin{equation}
\phi_{l}^{\sigma}(r)\propto e^{-r^{2}/2}r^{|c_{l\sigma}(\alpha)|}U\left(a_{l\sigma}^{s}(\alpha;\lambda^{2}),1+|c_{l\sigma}(\alpha)|,r^{2}\right)
\end{equation}
where $U(a,b,x)$ is \emph{Tricomi's hypergeometric function}. Near the origin they behave like $\approx r^{-|c_{l\sigma}(\alpha)|}$ for $\alpha\ne2l\sigma+1$, and as $\approx\log r$ for $\alpha=2l\sigma+1$. In the range of conical defects $\alpha\in(0,1]$, the only square-integrable solutions are for the zero-modes
\begin{equation}
\phi_{0}^{\sigma}(r)\propto e^{-r^{2}/2}r^{|c_{0\sigma}(\alpha)|}U\left(a_{0\sigma}^{s}(\alpha;\lambda^{2}),1+|c_{0}(\alpha)|,r^{2}\right).
\end{equation}
A quick computation shows that these behaviors lead to $||G_1^{s}\Phi||^{2}=\infty$, hence these solutions are not in the domain of the supercharge, which then has no complex eigenvalues at all. This shows that the deficiency subspaces $\mathcal{K}_{\pm}$ are trivial, hence we conclude that the supercharge is essentially self-adjoint in the range of conical defects.

In the range of conical surplusses $\alpha>1$, the solution $\phi_{l}^{\sigma}(r)$ is square-integrable if $|c_{l\sigma}(\alpha)|<1$, while one can directly see that $||G_1^{s}\Phi||<\infty$ if $-2<c_{l\sigma}(\alpha)<0$. Combining these two requirements we find that if 
\begin{equation}
-\frac{\alpha+1}{2}<l\sigma<\frac{\alpha-1}{2},
\end{equation}
the mode $\phi_{l}^{\sigma}$ induces an honest eigenstate of $G_1^{s}$ with eigenvalue $\lambda$, so populating the deficiency subspaces $\mathcal{K}_{\pm}$. Denoting $\kappa=\left\lfloor \frac{\alpha+1}{2}\right\rfloor $,  the dimensions of $\mathcal{K}_{\pm}$ are the same and equal to $4\kappa$ (or $4\kappa-2$ when $\alpha$ is an odd integer). We conclude that for conical surplusses the supercharge is not essentially self-adjoint, but it admits self-adjoint extensions.

\paragraph{Self-adjoint boundary conditions}

In the range of conical defects $\alpha\in(0,1]$ we showed that the supercharge $G_1^{s}$ is essentially self-adjoint. Its unique self-adjoint extension is precisely its adjoint $(G_1^{s})^{*}$, whose domain consists of wavefunctions satisfying the boundary  conditions \eqref{eq:bry_cond_adjoint_supercharge}.

In the range of conical surplusses $\alpha>1$, the deficiency subspaces $\mathcal{K}_{\pm}$ contain the states 
\begin{equation}\begin{aligned}
& (\pm i\Phi_{l}^{\sigma}+G_1^{s}\Phi_{l}^{\sigma}), \\
& \text{where }\Phi_{l}^{\sigma}(r,\theta)\propto e^{-r^{2}/2}r^{|c_{l\sigma}(\alpha)|}U\left(a_{l\sigma}^{s}(\alpha;-1),1+|c_{l\sigma}(\alpha)|,r^{2}\right)e^{il\theta/\alpha}|\sigma\rangle
\end{aligned}
\end{equation}
for every combination $l\sigma$ in the range $-\kappa,\cdots,-1,0,1,\cdots,\kappa-1$, corresponding to $-1<c_{l\sigma}(\alpha)<0$ (when $\alpha$ is an odd integer, the mode with $l\sigma=\kappa-1$ behaves logarithmically and has to be excluded). Notice that $\Phi_{l}^{\sigma}$ diverges like $r^{c_{l\sigma}(\alpha)}$ while $G_1^{s}\Phi_{l}^{\sigma}$ converges like $r^{c_{l\sigma}(\alpha)+1}$ near the origin. Since $\mathcal{K}_{\pm}$ are nonempty, for any unitary isomorphism  $\mathcal{U}:\mathcal{K}_{+i}\to\mathcal{K}_{-i}$ there is a non-minimal self-adjoint extension whose domain is 
\begin{equation}
\mathcal{D}_{\mathcal{U}}=\overline{\mathcal{V}}\oplus(1+\mathcal{U})\mathcal{K}_{+},
\end{equation}
where $\overline{\mathcal{V}}$ is the domain of the closure of $G_1^{s}$. By convention we take $\mathcal{U}\in U(4\kappa)$ to act as
\begin{equation}
\mathcal{U}\cdot(i\Phi_{l}^{\sigma}+G_1^{s}\Phi_{l}^{\sigma})=\sum_{l\sigma=-\kappa}^{\kappa-1}\mathcal{U}_{(\sigma l)'}^{(\sigma l)}(-i\Phi_{l'}^{\sigma'}+G_1^{s}\Phi_{l'}^{\sigma'}).
\end{equation}
Since elements of $\overline{\mathcal{V}}$ are regular at the conical singularity, any possible nontrivial behavior must come from elements in the summand $(1+\mathcal{U})\mathcal{K}_{+}$. In general, for any $\mathcal{U}\neq\mathds{1}$ some of the modes $\psi_{l}^{\sigma}$, for $l\sigma\in\{-\kappa,\cdots,\kappa-1\}$, of a wavefunction $\Psi\in\mathcal{D}_{\mathcal{U}}$ will be allowed to diverge at the origin like $r^{-|c_{l\sigma}(\alpha)|}$, so relaxing the boundary conditions of the essentially self-adjoint case.

The unique extension associated with $\mathcal{U}=\mathds{1}$ corresponds to regular boundary conditions for all the modes, because
\begin{equation}
(1+\mathcal{U})\cdot(i\Phi_{l}^{\sigma}+G_1^{s}\Phi_{l}^{\sigma})=(i\Phi_{l}^{\sigma}+G_1^{s}\Phi_{l}^{\sigma})+(-i\Phi_{l}^{\sigma}+G_1^{s}\Phi_{l}^{\sigma})=2G_1^{s}\Phi_{l}^{\sigma}
\end{equation}
which as we noted is regular at the origin. So $\mathcal{D}_{\mathds{1}}$ is in some sense the ``minimal'' domain of self-adjointness for the conical surplusses.

As another example, let us take $\mathcal{U}=\mathds{1}-2\delta_{\overline{l\sigma}}$. In the corresponding domain only the modes with $l\sigma=\overline{l\sigma}$ have nontrivial boundary conditions:
\begin{equation}
(1+\mathcal{U})\cdot(i\Phi_{l}^{\sigma}+G_1^{s}\Phi_{l}^{\sigma})=\begin{cases}
l\sigma=\overline{l\sigma}: & 2i\Phi_{\bar{l}}^{\bar{\sigma}}\approx r^{-|c_{\overline{l\sigma}}(\alpha)|}\\
l\sigma\neq\overline{l\sigma}: & 2G_1^{s}\Phi_{l}^{\sigma}\approx o(1).
\end{cases}
\end{equation}
In the case $\mathcal{U}=-\mathds{1}$ instead every mode $\psi_{l}^{\sigma}$ is allowed to diverge precisely as $r^{-|c_{l\sigma}(\alpha)|}$. When $\mathcal{U}$ is non-diagonal only some combinations of modes is allowed to diverge. For example let us assume that $\mathcal{U}$ mixes two labels $(l\sigma)_{1}$ and $(l\sigma)_{2}$ and acts as the identity on the other modes: the boundary conditions will be of the form
\begin{equation}
(1+\mathcal{U})\cdot(i\Phi_{l}^{\sigma}+G_1^{s}\Phi_{l}^{\sigma})\approx\begin{cases}
\# \Big( (\delta_{(\sigma l)_{1}}^{(\sigma l)}-\mathcal{U}_{(\sigma l)_{1}}^{(\sigma l)})
r^{-|c_{(l\sigma)_{1}}(\alpha)|}e^{il_{1}\theta/\alpha}|\sigma_{1}\rangle  & \\
\qquad +(\delta_{(\sigma l)_{2}}^{(\sigma l)}-\mathcal{U}_{(\sigma l)_{2}}^{(\sigma l)})r^{-|c_{(l\sigma)_{2}}(\alpha)|}e^{il_{2}\theta/\alpha}|\sigma_{2}\rangle \Big)
+ o(1) & l\sigma=(l\sigma)_{1},(l\sigma)_{2}\\
o(1) & \text{otherwise}
\end{cases}
\end{equation}
where only the overall normalization is unphysical while the relative normalization of the two terms in the first line is fixed and depends on $\mathcal{U}$.

\paragraph{BPS spectrum}

To find the complete spectrum of the supercharge, instead of solving the non diagonal equation  $G_1^{s}\tilde{\Phi}=\lambda\tilde{\Phi}$,  we can look at the Hamiltonian equation $\mathcal{H}_{s}\Phi=\lambda^{2}\Phi$ and then keep only the solutions $\Phi$ lying in some domain of self-adjointness for $G_1^{s}$. From $\Phi$ then we find the doublet of supercharge eigenstates $\tilde{\Phi}_{\pm}=(\pm|\lambda|\Phi+G_1^{s}\Phi)$ with eigenvalues $\pm|\lambda|$ whenever $\lambda\neq0$. BPS states instead  have eigenvalue $\lambda=0$: it can either happen that $G_1^{s}\Phi=0$, in which case we set $\tilde{\Phi}_{0}\equiv\Phi$, or if $G_1^{s}\Phi\neq0$ we set $\tilde{\Phi}_{0}\equiv G_1^{s}\Phi$.

Let us focus on the BPS spectrum. The Hamiltonian equation $\mathcal{H}_{s}\Phi=0$, after the same steps leading to \eqref{eq:Kummer_equation} becomes Kummer's equation with parameters $a_{l\sigma}^{s}(\alpha;0)=\frac{1}{2}(1+|c_{l\sigma}(\alpha)|+s\sigma(c_{l\sigma}(\alpha)+1))$ and $b_{l\sigma}(\alpha)=1+|c_{l\sigma}(\alpha)|$. Notice that the parameter $a_{l\sigma}^{s}(\alpha;0)$ is never a negative integer, so it follows that the only locally square-integrable eigenvalue which decays at $r\to\infty$ is
\begin{equation}
\Phi_{l}^{\sigma}(r,\theta)\propto e^{-r^{2}/2}r^{|c_{l\sigma}(\alpha)|}U\left(a_{l\sigma}^{s}(\alpha;0),1+|c_{l\sigma}(\alpha)|,r^{2}\right)e^{il\theta/\alpha}|\sigma\rangle
\end{equation}
for every $l,\sigma$. By direct computation one realizes that 
\begin{equation}
G_1^{s}\Phi_{l}^{-s}=0,\qquad G_1^{s}\Phi_{l}^{s}=\Phi_{l+s}^{-s},
\end{equation}
so the only independent BPS states for the supercharge $G_1^{s}$ are $\tilde{\Phi}_{l}^{(s)}\equiv\Phi_{l}^{\sigma=-s}$. Explicitly,
\begin{equation}
\tilde{\Phi}_{l}^{(s)}(r,\theta)\propto r^{c_{l(-s)}(\alpha)}e^{il\theta/\alpha}e^{-r^{2}/2}| -s \rangle.
\end{equation}

To fix the spectrum, we now have to restrict to those $\tilde{\Phi}_{l}^{(s)}$ which lie in some domain of self-adjointness for  $G_1^{s}$. They are square-integrable for $c_{l(-s)}(\alpha)>-1$, and moreover regular for $c_{l(-s)}(\alpha)\geq0$. In the range $\alpha\in(0,1]$ where $G_1^{s}$ is essentially self-adjoint, the acceptable states are those for $sl\leq0$ and they are all regular at the origin:  resorting to the complex coordinate $z=(\alpha re^{i\theta})^{1/\alpha}$ we find chiral and antichiral BPS states
\begin{equation}
\tilde{\Phi}_{|l|}^{(-)}\propto z^{|l|} |z|^{\frac{1-\alpha}{2}} e^{-K(|z|)} |+\rangle , \qquad \tilde{\Phi}_{|l|}^{(+)}\propto \bar{z}^{|l|} |z|^{\frac{1-\alpha}{2}} e^{-K(|z|)} |-\rangle \qquad |l|\geq0.
\end{equation}
Notice that for conical defects these all decay to zero, while in the plane case $\alpha=1$ there are two states with $l=0$ that are regular but finite at the origin.

For $\alpha>1$ there the situation complicates: if $\alpha$ is in the range $(2\kappa-1,2\kappa+1)$, there are $2\kappa$ singular states with $sl=-\kappa+1,\cdots,\kappa-1,\kappa$, and every other state with $sl\leq\kappa$ is regular (see Fig. \ref{fig:neg_pos_BPS_states}). All the states are allowed only in the maximally extended domain $\mathcal{D}_{\mathds{1}}$, where every $\tilde{\Phi}_{|l|}^{(s)}$ can diverge as $r^{-|c|}$ for $-1<c<0$ independently. The minimal domain $\mathcal{D}_{-\mathds{1}}$ instead removes all diverging BPS states, while all other domains allow only some linear combinations of the diverging ones. For example, if $\mathcal{U}=\mathds{1}-2\delta_{\overline{l}}$ then only the diverging BPS state with quantum number $l=\overline{l}$ will be included, etcetera.

\subsection{Basic notions about self-adjoint extensions} \label{app:basic_notions_selfadj}

To be self-contained, here we condense some theoretical notions that one needs to employ to study self-adjointness of the supercharges in
our models. These notions are standard in functional analysis and can be found for example in \cite{Reed:1975uy}.

Let $(\mathcal{H},\langle\cdot,\cdot\rangle)$ be a Hilbert space and $A:\mathcal{D}(A)\to\mathcal{H}$ an (unbounded) linear operator acting on a dense subspace $\mathcal{D}(A)\subset\mathcal{H}$ called its \emph{domain}. The operator $A$ is said to be \emph{symmetric} if $\langle A\phi,\psi\rangle=\langle\phi,A\psi\rangle$ for every $\phi,\psi\in\mathcal{D}(A)$.

For unbounded operators on infinite dimensional Hilbert spaces, the specification of domain is crucial. An \emph{extension} of $A$ is another linear operator $B:\mathcal{D}(B)\to\mathcal{H}$ which acts on a bigger domain $\mathcal{D}(B)\supset\mathcal{D}(A)$ and such that $B|_{\mathcal{D}(A)}=A$. The \emph{adjoint operator} $A^{*}:\mathcal{D}(A^{*})\to\mathcal{H}$ is defined in this way: its domain $\mathcal{D}(A^{*})$ constitutes of those elements $\psi\in\mathcal{H}$ such that there exists another element $\eta^{\psi}\in\mathcal{H}$ satisfying
\begin{equation}
\langle\eta^{\psi},\phi\rangle=\langle\psi,A\phi\rangle\qquad\forall\phi\in\mathcal{D}(A).
\end{equation}
Then one denotes the action $A^{*}\psi\equiv\eta^{\psi}$. If $A$ is symmetric, it is clear from the definition that $\mathcal{D} (A)\subset\mathcal{D}(A^{*})$ so $A^{*}$ is an extension of $A$ (in particular, it can have complex eigenvalues). A symmetric operator is called \emph{self-adjoint} if it is strictly equal to its adjoint, so  $\mathcal{D}(A)=\mathcal{D}(A^{*})$. Self-adjoint operators have real spectra. The double adjoint $(A^{*})^{*}\equiv\overline{A}$ of a symmetric, densely-defined operator is called its \emph{closure}.\footnote{The closure $\overline{A}$ of any densely defined operator $A$ is actually a notion independent of its adjoint, but we will not need its general definition.
%It is defined (when it exists) as the continuous extension of $A$ whose graph is the closure in $\mathcal{H}\times\mathcal{H}$ of the graph of $A$. Any symmetric operator is closable, in that $\overline{A}$ exists and coincides with the double adjoint $\overline{A}=A^{**}$. Sometimes the domain of $\overline{A}$ is easier to compute following this direct definition. A useful result is that $\mathcal{D}(\overline{A})$ can be found as the closure $\overline{\mathcal{D}(A)}^{||\cdot||_{A}}$ in $\mathcal{H}$ with respect to the norm $||\psi||_{A}:=\sqrt{||\psi||^{2}+||A\psi||^{2}}$.
} It is a symmetric extension of $A$ which satisfies
\begin{equation}
\mathcal{D}(A)\subset\mathcal{D}(\overline{A})\subset\mathcal{D}(A^{*}).
\end{equation}

When we are given a symmetric, densely-defined operator, we can look for symmetric extensions or self-adjoint extensions of it. We mentioned that the closure $\overline{A}$ is a symmetric extension but generically it is not self-adjoint. The adjoint $A^{*}$ is instead generically not even symmetric. When it happens to be symmetric, then it is easy to realize that it must be also self-adjoint, so $A^{*}=\overline{A}$. In this case $A^{*}$ is the \emph{unique} self-adjoint extension of $A$, and $A$ is called \emph{essentially self-adjoint}. 

We are also interested in cases when $A^{*}$ is not symmetric. Then $A$ can have many self-adjoint extensions, each of them having a different domain. The strategy to classify these extensions is due to a series of theorems by Von Neumann. One defines the \emph{deficiency subspaces} 
\begin{equation}
\mathcal{K}_{\pm}:=\ker(A^{*}\mp i)\qquad\subset\mathcal{D}(A^{*}),
\end{equation}
which are the eigenspaces of $A^{*}$ corresponding to the complex eigenvalues $\pm i$. It turns out that: 
\begin{itemize}
\item If $\dim\mathcal{K}_{+}=\dim\mathcal{K}_{+}=0$, then $A^{*}$ has no complex eigenvalues and it is actually self-adjoint. So $A$ is essentially self-adjoint. 
\item If $\dim\mathcal{K}_{+}\neq\dim\mathcal{K}_{+}$, then $A$ has no self-adjoint extensions.
\item If $\dim\mathcal{K}_{+}=\dim\mathcal{K}_{+}=:n>0$, then $A$ has a $U(n)$-worth of self-adjoint extensions. In particular, to every unitary isomorphism $\mathcal{U}:\mathcal{K}_{+}\to\mathcal{K}_{-}$ it corresponds the domain of self-adjointness
\begin{equation}
\mathcal{D}_{\mathcal{U}}(A):=\mathcal{D}(\overline{A})\oplus(1+\mathcal{U})\mathcal{K}_{+}.
\end{equation}
\end{itemize}
Often in physics applications one is handed a ``formal'' hermitian differential operator $A$ acting on a Hilbert space $\mathcal{H}$, without a proper specification of domain of $A$. In our examples, this is a supercharge acting on the space of $L^{2}$ spinors over (possibly open) manifolds. One task is then to understand if $A$ is self-adjoint, i.e. a physical observable whose spectrum is real. The strategy to do this is:
\begin{enumerate}
\item Make an initial ``conservative'' choice for the domain $\mathcal{D}(A)$, which has to be dense in $\mathcal{H}$, where the action of $A$ is simple and where it is easy to establish that $A$ is symmetric.
\item When possible, study the domains of $A^{*}$ and $\overline{A}$. If it turns out that $A^{*}=\overline{A}$ is self-adjoint, then one concludes that $A$ is essentially self-adjoint. The domain of self-adjointness is $\mathcal{D}(A^{*})$.
\item Study the deficiency subspaces $\mathcal{K}_{\pm}$. When $\dim\mathcal{K}_{+}=\dim\mathcal{K}_{-}>0$, classify the domains of self-adjointness $\mathcal{D}_{\mathcal{U}}$.
\end{enumerate}

\section{The Dirac index} \label{App:Dirac_index}
Here we review and motivate a few important points about the localization formulas \eqref{ABformisolated},\eqref{ABformnonisolated}. As we noticed in Section \ref{sec:geom_interpretation}, the unrefined index on a smooth target space $\mathcal{M}$ formally reproduces the index of a Dirac operator coupled to a $U(1)$ background gauge field. We denote by $L$ the line bundle associated with this background field, so that spinors take value in the spinor bundle twisted by $L$. The famous index theorem by Atiyah and Singer gives the formal result \cite{AtiyahSinger_elliptic_III1968,Eguchi:1980jx}
\be
\Omega=\int_{\mathcal{M}}\mathrm{ch}(L)\wedge\hat{A}(T\mathcal{M}),
\ee 
where $\mathrm{ch}(L)$ is the Chern character of the line bundle $L$ and $\hat{A}(T\mathcal{M})$ is the "A-hat" genus of the tangent bundle.\footnote{We will not dwell here on the fact that our target $\cal M$ is non-compact, and consider this as a formal result. This is justified due to the discussion in Section \ref{sec:smooth_and_loc}, see also \cite{Raeymaekers:2024usy}.}
Notice that the index depends only on the principal symbol of the Dirac operator and thus not on the specific Riemannian connection (and torsion) used to define it, as it is evidenced by the appearance of characteristic classes on the rhs of the above formula. Starting from 
%These characteristic classes are topological and can be computed using 
any curvature tensor $F\in\Omega^{2}(\mathcal{M})$ for the bundle $L$ and $R\in\Omega^{2}(\mathcal{M};\mathrm{End}(T\mathcal{M}))$ for the tangent bundle, they can be computed via the formulas
\be
\mathrm{ch}(L)=e^{-F/2\pi},\qquad\hat{A}(T\mathcal{M})=\sqrt{\det\left(\frac{R/2}{\sinh(R/2)}\right)}. \label{eq:char_classes}
\ee
%\commentP{might have to normalize every curvature by $i/2\pi$. Especially the immaginary units have to be chosen based on F and R being real or imaginary. To check}

Assume now that there is a torus $T=(U(1))^{l}$ group action on the manifold $\mathcal{M}$, generated infinitesimally by a set of vector fields $\{j_{I}\in\Gamma(T\mathcal{M})\}_{I=1,\cdots,l}$ spanning a commutative Lie algebra $\mathfrak{t}$.\footnote{Here, to uniformize notation, we consider at the same time all the isometries without distinguishing the $R$-symmetry.} Then the index formula can be refined using tools from $T$-equivariant
cohomology as follows. Let us fix a group element $q=(e^{i\nu^{1}},\cdots,e^{i\nu^{l}})\in T$.\footnote{Equivalently the set of chemical potentials $\{\nu^{I}\}$ can be thought as the basis of the Lie coalgebra $\mathfrak{t}^{*}$, dual to the generators $\{j_{I}\}$.} The ``character-valued index'' or ``Lefschetz number'' \eqref{GindDirac} can be obtained by the formula %\commentP{ref and/or quote localization argument} 
%\commentP{to be precise should assume that the action lifts to the bundles and that the Dirac operator is $T$-equivariant, which in particular is true thanks to $\mathcal{L}_{j_{I}}C=0$, invariance of the metric and of the background field. Leave implicit?}
\begin{equation} \label{eq:char_valued_index_theorem}
\Omega[\nu^{I}]=\int_{\mathcal{M}}\mathrm{ch}_{T}(L)\wedge\hat{A}_{T}(T\mathcal{M}),
\end{equation}
where now we integrate the $T$-equivariant cohomology classes $\mathrm{ch}_{T}(L),\hat{A}_{T}(T\mathcal{M})\in H_{T}^{\bullet}(\mathcal{M})$,\footnote{We skip the details about the construction of $H_{T}^{\bullet}(\mathcal{M})$, as they are not too important for us here. For reference, see e.g.\ \cite{Tu_book_eq_coho2020}. In short, it can be computed analogously to the well-known de Rham cohomology, where instead of differential forms one starts from $T$-equivariant forms on $\mathcal{M}$, and instead of the de Rham differential $d$ one uses a certain equivariant extension of it denoted $d^{T}$, see below in the text for more details.} which are implicitly polynomials in the chemical potentials $\{\nu^{I}\}$. Picking now $T$-invariant curvature tensors $F$ and $R$ for the line and tangent bundle respectively, $\rm{ch}_T$ and $\hat{A}_{T}$ can be computed replacing in \eqref{eq:char_classes} the so-called ``$T$-equivariant extensions''
\begin{equation}
F^{T}=F- 2\pi i \nu^{I}\mu_{I}^{L},\qquad R^{T}=R - i \nu^{I}\mu_{I}^{T\mathcal{M}}.
\end{equation}
In these expressions, the \emph{moment maps} $\mu^{L}$ and $\mu^{T\mathcal{M}}$ are defined by the ``Hamiltonian equations''
\begin{equation}
d\mu_{I}^{L}= \iota_{j_{I}}F / 2\pi i,\qquad i \nabla\mu_{I}^{T\mathcal{M}}=\iota_{j_{I}}R,
\end{equation}
plus the $T$-equivariant constraint $\mathcal{L}_{j_{I}}\mu_{j_{J}}=0$. In particular notice that $\mu_{I}^{L}$ (resp. $\mu^{T\mathcal{M}})$ are global smooth functions (sections of the bundle $\mathrm{End}(T\mathcal{M})$) defined up to an additive locally constant (covariantly constant) term. One can check that a good canonical choice for the moment map of the tangent bundle is $\mu_{I}^{T\mathcal{M}}=-\nabla j_{I}$. 
%\commentP{stress about existence problem? They do exist but the formula $\iota_{j_{I}}A$ is only local and valid up to an additional term. In examples better to solve for the Hamiltonian equation directly}

In presence of a torus symmetry, the character-valued index formula can be substantially simplified thanks to a finite dimensional localization theorem due to Berline-Vergne and Atyiah-Bott \cite{berline1982classes, Atiyah:1984px}. It states the following: Consider a $T$-equivariant form of top-degree 
\begin{equation}
\alpha=\alpha^{(D)}+\nu^{I}\alpha_{I}^{(D-2)}+\nu^{I}\nu^{J}\alpha_{IJ}^{(D-4)}+\cdots+\nu^{I_{1}}\cdots\nu^{I_{D/2}}\alpha_{I_{1}\cdots I_{D/2}}^{(0)},
\end{equation}
which is a polynomial in the $\nu^{I}$ (formally considered in degree 2) whose coefficients $\alpha_{I_{1}\cdots I_{k}}^{(D-2k)}\in\Omega^{D-2k}(\mathcal{M})^{T}$
are $T$-invariant forms on $\mathcal{M}$. If $\alpha$ is closed with respect to the $T$-equivariant differential 
\begin{equation}
d^{T}=d+\nu^{I}\iota_{j_{I}},
\end{equation}
then its integral over $\mathcal{M}$ localizes to the submanifold $\mathcal{M}^{T}\xhookrightarrow{i}\mathcal{M}$ of fixed points under the torus action, corresponding to the common zero locus of the vector fields $\lbrace j_I\rbrace$. In formulae,
\begin{equation}
\int_{\mathcal{M}}\alpha=\int_{\mathcal{M}^{T}}\frac{i^{*}\alpha}{e_{T}(N)}
\end{equation}
where $e_{T}(N)$ is the $T$-equivariant Euler class of the normal bundle $N$ of $\mathcal{M}^{T}$ in $\mathcal{M}$. The latter can be computed from the formula
\begin{equation}
e_{T}(N)=\mathrm{Pf}\left(R^{T}(N)\right),\qquad R^{T}(N)=R(N)- i \nu^{I}\mu_{I}^{N},
\end{equation}
where $R^{T}(N)$ is the equivariant extension of the curvature of the normal bundle %\commentP{might be off by factors of $2\pi$ in the normalization of the curvature, but I think this is consistent with the def of the other classes above} 
and $\mathrm{Pf}$ denotes the pfaffian, which for antisymmetric matrices coincides with the square root of the determinant. To make sense of the integrand on the rhs, one should think of expanding it in power series of differential
forms and then pick the correct term according to the dimension of the submanifold $\mathcal{M}^{T}$. Notice that the equality holds as polynomials in the $\nu^{I}$'s.

Notice that our equivariant extensions $F^{T}$ and $R^{T}$ are precisely constructed so that their characteristic classes are closed with respect to $d^{T}$.\footnote{While the abelian gauge field $F^{T}$ itself is closed in $d^{T}$, $R^{T}$ is closed with respect to a covariantized version of $d^{T}$. But every ad-invariant polynomial in the curvature is then closed with respect to $d^{T}$.} This means that we can directly apply the ABBV localization theorem to \eqref{eq:char_valued_index_theorem}, obtaining
\begin{equation}
\Omega[\nu^{I}]=\int_{\mathcal{M}^{T}}\frac{i^{*}\left(\mathrm{ch}_{T}(L)\wedge\hat{A}_{T}(T\mathcal{M})\right)}{e_{T}(N)}.
\end{equation}

Let us notice that the submanifold ${\cal M}^T$ is always of even codimension. Since we are working with an even-dimensional target $\cal M$, this means that index can in some cases localize to a sum over isolated fixed point in the case where $\dim{\cal M}^T = 0$. For computability purposes, we will review below how to simplify the localization formula separately for the cases of isolated and non-isolated fixed points.

\subsection{Isolated fixed points}

For isolated fixed points $\mathcal{M}^{T}$ is zero-dimensional and the normal bundle is the restriction of the full tangent bundle to $\mathcal{M}^{T}$. Accordingly, only the 0-form part of the various characteristic classes survive in the localization formula, which in this case simplifies to
\begin{align}
\Omega[\nu^{I}] & =\sum_{p\in\mathcal{M}^{T}}\left.\frac{e^{i \nu^{I}\mu_{I}^{L}}}{\mathrm{Pf}(i \nu^{I}\nabla j_{I})}\sqrt{\det\left(\frac{i \nu^{I}\nabla j_{I}}{2i \sin(\nu^{I}\nabla j_{I}/2)}\right)}\right|_{p}\\
 & =\sum_{p\in\mathcal{M}^{T}}e^{i \nu^{I}\mu_{I}^{L}(p)}\left.\det\left(2i \sin(\nu^{I}\nabla j_{I}/2)\right)^{-1/2}\right|_{p}
\end{align}
where we see cancelations between the Pfaffian at the denominator and the numerator of the A-hat genus. 

To compute the determinant, we notice that at every fixed point $p\in\mathcal{M}^{T}$ the corresponding tangent space decomposes into irreducible representations of the $T$-action,\footnote{Notice that in the irreducible composition of $T{\cal M}^T|_p$ we cannot find the trivial representation: if a tangent vector is fixed by the torus action, then its whole integral curve starting from $p$ would be fixed, contradicting the assumption of isolated fixed points. This shows that isolated fixed points can be found only on even-dimensional manifolds.}
\begin{equation} \label{eq:tangent_space_decomposition_torus_action}
T\mathcal{M}|_{p}\cong\sum_{\mu=1}^{D/2}V_{l_{\mu}(p)},\qquad V_{l_{\mu}(p)}\cong\mathbb{R}^{2}
\end{equation}
for some ``exponents'' or ``weights'' $l_{\mu}(p)\in\mathfrak{t}^{*}$. This means that there exist suitable Daboux-like coordinates $\{x_{(p)}^{\mu},y_{(p)}^{\mu}\}^{\mu=1,\cdots,D/2}$ around $p$ where the Killing vectors $j_{I}$ can be expressed as 
\begin{equation}
j_{I}|_{p}=\sum_{\mu=1}^{n}l_{I,\mu}(p)(y_{(p)}^{\mu}\partial_{x_{(p)}^{\mu}}-x_{(p)}^{\mu}\partial_{y_{(p)}^{\mu}})
\end{equation}
In these coordinates the skew-symmetric matrix $\nabla j_{I}$ at any fixed point is block-diagonal with skew-eigenvalues $\{l_{I,\mu}(p)\}$.

With this notation, the index formula for isolated fixed points simplifies to
\begin{equation}
\Omega[\nu^{I}]=\sum_{p\in\mathcal{M}^{T}}e^{i\nu^{I}\mu_{I}^{L}(p)}\prod_{\mu=1}^{n}\frac{1}{2i \sin(\nu^{I}l_{I,\mu}(p)/2)}.
\end{equation}

\subsection{Non isolated fixed points}

For non-isolated fixed points the derivation is more involved. Near the fixed point locus we can split the tangent bundle as $T\mathcal{M}|_{\sim\mathcal{M}^{T}}=T\mathcal{M}^{T}\oplus N$, and pick a $T$-invariant connection whose curvature tensor also block-diagonalizes like
\begin{equation} \label{eq:curvature_matrix_dec}
R=\left(\begin{matrix}R^{t} & 0\\
0 & R^{n}
\end{matrix}\right),
\end{equation}
where ``$t$'' and ``$n$'' denote tangent and normal directions in some suitable adapted coordinates.\footnote{Notice that the normal bundle is defined as the quotient $N\cong T\mathcal{M}/T\mathcal{M}^{T}$ so the tangent bundle does not split canonically. A way to chose a split is to employ a $T$-invariant metric, which will tautologically block-diagonalize as $G=G^{t}\oplus G^{n}$ in a tubular neighborhood of $\mathcal{M}^{T}$. One can see that, in coordinates adapted
to this splitting, the Christoffel symbols and hence also the Riemannian curvature tensor will block-diagonalize.} Since the Killing vector fields $j_{I}$ vanish on $\mathcal{M}^{T}$, the moment map will decompose schematically as 
\begin{equation} \label{eq:moment_map_matrix_dec}
\nabla j_{I}=\left(\begin{matrix}0 & 0\\
\partial_{n}j_{I}^{t} & \partial_{n}j_{I}^{n}
\end{matrix}\right),
\end{equation}
so the equivariant extension $R^{T}$ of the curvature in these coordinates will be a lower triangular matrix of polyforms. The same is true for every power of $R^{T}$, so we conclude that the equivariant $\hat{A}$-genus will factorize as $\hat{A}_{T}(T\mathcal{M})|_{\mathcal{M}^{T}}=\hat{A}_{T}(N)\hat{A} (T\mathcal{M}^{T})$, where by the form of \eqref{eq:curvature_matrix_dec} and \eqref{eq:moment_map_matrix_dec} we notice that only the factor corresponding to the normal bundle is $T$-equivariantized
(the moment map along the tangent directions vanishes). The combination $e_{T}(N)^{-1}\hat{A}_{T}(N)$ will see partial cancellation between the Euler class and the numerator of the $\hat{A}$-genus of the normal bundle, similarly to the case of isolated fixed-points. This leads to the formula
\begin{equation}
\Omega[\nu^{I}]=\int_{\mathcal{M}^{T}}\mathrm{ch}_{T}(L)\wedge\hat{A}(T\mathcal{M}^{T})\wedge\det\left(2 \sinh(R^{n} + i \nu^{I}\partial_{n}j_{I}^{n}/2)\right)^{-1/2}.
\end{equation}

As a last simplification, we can express the determinants in terms of the skew-eigenvalues of the curvature tensors and of the moment maps. For this we need to make the extra assumption that the $T$-invariant curvature along the normal bundle $R^{n}$ is skew-diagonalized in the same basis that realizes the weight decomposition\footnote{By a similar argument as in the case of isolated fixed points, the normal bundle can contain only nontrivial representations with respect to the torus action and hence it is of even rank.}
\begin{equation}
N|_{p}\cong\sum_{\mu=1}^{\mathrm{rank}(N)/2}N_{l_{\mu}(p)},\qquad N_{l_{\mu}(p)}\cong\mathbb{R}^{2},\qquad\forall p\in\mathcal{M}^{T}
\end{equation}
in eigenspaces with respect to the $T$-action, analogous to \eqref{eq:tangent_space_decomposition_torus_action}. That is, we assume that we can express simultaneously 
%\commentP{Witten comments on this. Is it really an extra assumption?}
\begin{equation}
\partial_{n}j_{I}^{n}|_{p}\sim\left(\begin{matrix}\ddots\\
 &  & -l_{I,\mu}(p)\\
 & l_{I,\mu}(p)\\
 &  &  & \ddots
\end{matrix}\right),\quad R^{n}|_{p}\sim\left(\begin{matrix}\ddots\\
 &  & -x_{\mu}^{n}(p)\\
 & x_{\mu}^{n}(p)\\
 &  &  & \ddots
\end{matrix}\right)\qquad\forall p\in\mathcal{M}^{T}.
\end{equation}
The curvature $R^{t}$ will be also skew-diagonalized in some frame, with skew-eigenvalues $\{x_{r}^{t}(p)\}_{r=1,\cdots,\dim\mathcal{M}^{T}/2}$ but this poses no extra assumption since its determinant is computed independently. The elements $x_{\mu}^{n}$ are identified with the first Chern classes $[x_{\mu}^{n}]=c_{1}(N_{l_{\mu}})$ of the line bundles decomposing $N$, as well as $x_{r}^{t}$ being the first Chern classes of $T\mathcal{M}^{T}$ according to the splitting principle.

Under the above assumption and notation, we can write the character-valued index formula concretely as
\begin{equation}
\label{eq:loc_formula_nonisolated}
\Omega[\nu^{I}]=\int_{\mathcal{M}^{T}}e^{c_1(F) + i \nu^{I}\mu_{I}^{L}} \prod_{r=1}^{\dim\mathcal{M}^{T}/2}\frac{x_{r}^{t}}{2\sinh(x_{r}^{t}/2)}\prod_{\mu=1}^{\mathrm{rank}(N)/2}\frac{1}{2\sinh((x_{\mu}^{n} + i\nu^{I}l_{I,\mu})/2)}.
\end{equation}

\section{The conifold}\label{Appcon}

As an  illustrative example where the target space  is neither hyperk\"ahler nor an orbifold, let us compute the superconformal index on the  conifold. The conifold is a six-dimensional Ricci-flat  K\"ahler cone (hence $C_{ABC}=0$) whose base $\calb = T_{1,1}$ is topologically $S^3 \times S^2$. We refer to \cite{Candelas:1989js,PandoZayas:2000ctr} for more details on the conifold and its resolutions. The metric on the base is Sasaki-Einstein and given by
\be 
ds^2_{T_{1,1}} = {1\over 9} (d\psi + \cos \theta_1 d\f_1 + \cos \theta_2 d\f_2)^2
+{1 \over 6}(d\theta_1^2 + \sin^2 \theta_1 d \f_1^2+ d\theta_2^2 + \sin^2 \theta_2 d \f_2^2)
\ee
where $\psi \in [0, 4 \p)$ is a coordinate on the Hopf fiber. 
Complex coordinates can be chosen as follows
\bea
z^1 &=& R e^{- {i\over 2} (\psi + \f_1 + \f_2)}\cos {\theta_1\over 2}\cos {\theta_2\over 2}\nonu
z^2 &=& R e^{- {i\over 2} (\psi - \f_1 + \f_2)}\sin {\theta_1\over 2}\cos {\theta_2\over 2}\nonu
z^3 &=& e^{i \f_2} \tan {\theta_2 \over 2}\label{complcon}
\eea
where $R$ is related to the canonical radial coordinate $r$ in (\ref{conecoords}) as
\be 
R = \left( \sqrt{2\over 3} r\right)^{3 \over 2}.
\ee
 Computing the Reeb vector $\r$ one finds
 \be 
 \r = - {3i\over 2} \left( z^1 \pa_{z^1} 
 +  z^2 \pa_{z^2}  \right) + {\rm c.c.}= 3 \pa_\psi \label{reebconifold}
 \ee 
 In addition, we have two further real-holomorphic Killing vectors
 \be 
 j_1 = \pa_{\f_1}, \qquad  j_2 = \pa_{\f_2},
 \ee
 which satisfy the conditions (\ref{extrasymmconds}) to provide further refinements to the index.
We can also turn on a background  gauge field in this model while preserving superconformal symmetry. Indeed, if we take the $m$-monopole gauge potential
 \be 
 A = {m \over 2} (\pm 1 - \cos \theta_2)d\f_2,\label{Acon}
 \ee
 we see that it preserves the conditions (\ref{N2conds},\ref{Rconds}) for superconformal invariance since it is of type $(1,1)$ and satisfies $i\r F=0$.

 Let us breafly comment on the ambiguity of the smoothening procedure in this example. Due to the absence of torsion, the standard criterion \eqref{Choucrit} diagnoses the possible ambiguities in the BPS spectrum when working near the conical singularity, at least if we restrict to vanishing background gauge field. Combining this with the fact that the conifold is Ricci-flat, Theorem 1.3 in \cite{Chou1989} tells us that the normalizable BPS spectrum is unambiguous. We therefore expect the index to be reproduced by the smoothening procedure.\footnote{Still, it would be interesting to analyze if the monopoles \eqref{Acon} can qualitatively change the game.}

 There are two ways to smoothen out the conical singularity while preserving homogeneity and K\"ahlerity of the cone: the so-called small resolution and the deformation of the conifold. In the first, the $S^2$ remains of finite  size while the $S^3 $ shrinks to zero size  smoothly at the tip, while for the deformation the roles of the $S^2$ and $S^3$ are reversed. In order to apply  fixed-point formulas we want  the Reeb vector to still have a  fixed locus in the smoothened manifold, so we restrict our attention to the small resolution. 
 
 The explicit K\"ahler metric on the resolved conifold is \cite{PandoZayas:2000ctr}
 \bea
 ds^2_{\rm reg} &=& \g' d R^2 + { R^2 \g' \over 4} (d\psi + \cos \theta_1 d\f_1 + \cos \theta_2 d\f_2)^2\\ && + {\g \over 4} (d\theta_1^2 + \sin^2 \theta_1 d \f_1^2) +\left(  {\g \over 4} + a^2 \right)( d\theta_2^2 + \sin^2 \theta_2 d \f_2^2))
 \eea
 where $\g = \g (R^2)$ is a function of $R^2 $ only and $\g' \equiv \pa \g / \pa R^2$. The explicit  expression is
 \be 
 \g (R^2) = 2 a^2 \left( -1 + { 2 a^2 \over N^{1\over 3}} + {N^{1\over 3}\over 2 a^2} \right), \qquad N(R^2) = \half (R^4 - 16a^6 + \sqrt{ R^8 - 32 a^6 R^4})
 \ee
 (with a suitable branch choice when the argument of the root becomes negative).
 Here, $a$ is a resolution parameter such that the singular conifold metric is recovered as $a \to 0$. 
  Note that the conifold metric is also
 approached as $R \to \infty$ as per our general requirements.
 Complex coordinates $z^m$  adapted to the complex structure  are still given by (\ref{complcon}). These cover only one patch of the geometry excluding the south pole of the $S^2$. To fully cover the manifold we need a second patch with coordinates $u^m$, related to the $z$-coordinates on the overlap region as \cite{Candelas:1989js}
 \be
 ( z^1 , z^2, z^3 ) = \left( - u^2 u^3, - u^1 u^3, {1\over u^3}\right)
 \ee
 These transition functions reflect the fact that the resolved conifold is the total space of the bundle $\calo(-1) \oplus \calo(-1)$ over $S^2$.
 The vectors $\r^{\rm reg} = 3\pa_\psi, j_1^{\rm reg}= \pa_{\f_1}$ and $j_2^{\rm reg}= \pa_{\f_2} $ are  still  real-holomorphic Killing vectors.  
 For later reference we give their  expression  in both coordinate patches:
 \bea 
 \r^{\rm reg} &=&  - {3i\over 2} \left( z^1 \pa_{z^1}  
 +  z^2 \pa_{z^2}   \right) + {\rm c.c.}
 =  - {3i\over 2} \left( u^1 \pa_{u^1}
 +  u^2 \pa_{u^2} \right) + {\rm c.c.} \\
j_1^{\rm reg} &=& - {i\over 2} \left( z^1 \pa_{z^1}- z^2 \pa_{z^2} \right)  + {\rm c.c.} 
  =  {i\over 2} \left( u^1 \pa_{u^1}- u^2 \pa_{u^2} \right)  + {\rm c.c.} \\
 j_2^{\rm reg} &=& - {i\over 2} \left( z^1 \pa_{z^1}+  z^2 \pa_{z^2} \right) + i   z^3 \pa_{z^3}   + {\rm c.c.} 
  =  {i\over 2} \left( u^1 \pa_{u^1}
 +  u^2 \pa_{u^2} \right) - i  u^3 \pa_{u^3}  + {\rm c.c.}\qquad
 \label{KVscon}
 \eea
We take the background gauge field to be still of the form
(\ref{Acon}). From the definition (\ref{extrasymmconds}) we see that there is a nontrivial moment map associated to $j_2$:
\be 
\m_2^{\rm reg} = {m\over 2} \cos \theta_2.
\ee
Due to these properties we can define a refined   index on the resolved model   as
 \be 
 \O_\pm^{\rm reg}  [ q, x_1,x_2] = \tr (-1)^F  e^{- \b \calh_\pm}q^{ J} x_1^{J_1} x_2^{J_2}.\label{refindcon}
 \ee

\subsection{Unrefined index}
Let us first compute the unrefined index (\ref{refindcon}) with $x_1=x_2=1$.
  This localizes on the vanishing locus of    $\r^{\rm reg}$, i.e. $R=0$. From the behaviour of $\g$,
 \be 
 \g \sim {R^2\over \sqrt{6} a} + \calo (R^4),
 \ee
we see that this locus is a round two-sphere   of radius $a$. Therefore we need the index formula (\ref{ABformnonisolated}) for non-isolated fixed points. The 
normal bundle splits in two copies of $\calo(-1)$, which according to  each carry charge $l=3/2$ under the $\r^{\rm reg}$-action.
Letting $x$ be the first Chern class of $\calo (-1)$, the equivariant index formula then reduces to   \bea 
\O^{\rm reg} [e^{i\theta}] &=& \int_{S^2}   e^{c_1 (F)} \left( 2 \sinh  {x + {3\over 2} i  \theta\over 2}\right)^{-2}\\
&=& {q^{ 3 \over 2}\left(1+m + (1-m) q^{ 3 \over 2}\right) \over \left(1- q^{ 3 \over 2}\right)^3},\label{conindunref}
\eea
where in the last line we used $\int _{S^2 } c_1(F) = - m, \int _{S^2 } x =-1$.
%One should remark that, for $m=0$, this agrees with the following limit of Dorey ea's result (5.8) in \cite{Dorey:2019kaf}
%\be 
%q^{-{ 3 \over 2}} \O[ q] = - \lim_{\tilde y \to 0} \tilde y^{3\over 2}
%\calz [ \t = q^{c_J}, z_i = 1]
%\ee
%\comment{Explain why this particular limit.}

\subsection{Refined index}
We now compute the refined index  (\ref{refindcon}).
Taking $q = e^{i \theta}, x_{1,2} = e^{i \m^{1,2}}$, it  localizes on the vanishing locus of the generator
\be 
j_{\theta, \m^1, \m^2} =   \theta \r^{\rm reg} +  \m^1 j_1^{\rm reg} +  \m^2 j_2^{\rm reg}.
\ee
When $\m^2 \neq 0$, it is straightforward to see that there are two isolated fixed points at  
$z^i =0$ resp. $u^i =0$, i.e.  at the north and south poles of the two-sphere at $R=0$. From (\ref{KVscon})  we read of the exponents of the various $u(1)$  actions and values of the nontrivial moment map $\m_{j_2^{\rm reg}}$  at these fixed points:
%\bea 
%\call_{j_{\theta, \m^1, \m^2}}&=& {\rm diag} \left[i \left({3 \over 2} \theta   + {\m^1\over 2}+ {\m^2\over 2},{3 \over 2} \theta   - {\m^1\over 2}+ {\m^2\over 2}, - \m^2 \right)\right] \qquad {\rm at\ } z^m =0\\
%&=& {\rm diag} \left[i \left({3 \over 2} \theta   - { \m^1\over 2}- {\m^2\over 2},{3 \over 2} \theta   + {\m^1\over 2}- {\m^2\over 2},  \m^2 \right)\right] \qquad {\rm at\ } u^m =0
%\eea
\begin{align}
{\rm at\ } z^m =&0: & l_{\r^{\rm reg}} =& \left({3\over 2},{3\over 2},0\right) , & l_{j_1^{\rm reg}} =&\left({1\over 2}, -{1\over 2},0\right),&  l_{j_2^{\rm reg}} =&\left({1\over 2}, {1\over 2},-1\right) &\m_{j_2^{\rm reg}} =& {m\over 2} \nonu
{\rm at\ } u^m =&0: & l_{\r^{\rm reg}} =& \left({3\over 2},{3\over 2},0\right) , & l_{j_1^{\rm reg}} =&\left(-{1\over 2}, {1\over 2},0\right),&  l_{j_2^{\rm reg}} =&\left(-{1\over 2}, -{1\over 2},1\right) &\m_{j_2^{\rm reg}} =& - {m\over 2}
\end{align}
Applying the Atiyah-Bott formula (\ref{ABformisolated})  one obtains,
{
\small
\be
\O^{\rm reg} [ q, x_1,x_2] = {q^{{3 \over 2} } x_2^{-{m\over 2}}\over 1- x_2}\left( {1\over \left( 1-  q^{ {3 \over 2}} \left( {x_1  x_2 }\right)^{-\half} \right) \left( 1-  q^{ {3 \over 2}} \left( {x_1 \over x_2 }\right)^\half \right)} - { x_2^{m+1} \over  \left( 1-  q^{ {3 \over 2}} \left( {x_1  x_2 }\right)^{\half} \right) \left( 1-  q^{ {3 \over 2}} \left( {x_2 \over x_1 }\right)^\half \right) }\right)
\ee
}One checks that, in the limit $x_{1,2}\to 1$, this reduces to the unrefined result (\ref{conindunref}).
For $m=0$ this the expression can be rewritten as\footnote{According to \eqref{ABrel} and to (4.17) in \cite{Dorey:2019kaf}, $q^{-3/2}\Omega^{\rm reg}[q,x_1,x_2]|_{m=0}$ matches with the Hilbert series for the conifold. Indeed, one can perform the substitution $(q,x_1,x_2)\to (t_3^{1/3}, t_1 t_2/t_3, t_1/t_2)$ and compare with e.g. (A.5) in \cite{Forcella:2008bb}.}
\be 
 \O^{\rm reg} [ q,x_1,x_2]_{| m=0} = { q^{{3 \over 2}}\left( 1 -  q^{ 3 }   \right)\over \left( 1-  q^{{3 \over 2}} \left( {x_1  x_2 }\right)^\half \right) \left( 1-  q^{{3 \over 2}} \left( {x_1  x_2 }\right)^{-\half} \right) \left( 1-  q^{{3 \over 2}} \left( {x_2 \over x_1 }\right)^\half \right) \left( 1-  q^{ {3 \over 2}} \left( {x_1 \over x_2 }\right)^\half \right) }.\label{refindconm0}
\ee
%which reduces to the unrefined result with $m=0$ upon setting $x_1=x_2=1$ and is equivalent to (5.8) in \cite{Dorey:2019kaf}. 

\end{appendix}                                                                                                                                                                       
\bibliographystyle{ytphys}
\bibliography{refssing}
\end{document}